\documentclass[final,5p,times,twocolumn]{elsarticle}
\usepackage{hyperref,textcomp}

\newcommand*{\megsign}{\mu^+ \to \mathrm{e}^+ \gamma} 

\begin{document}

\begin{frontmatter}

\title{Large-Area MPPC with Enhanced VUV Sensitivity for Liquid Xenon Scintillation Detector}




\author[address_icepp]{K.~Ieki}
\author[address_icepp]{T.~Iwamoto}
\author[address_icepp]{D.~Kaneko\fnref{now}}
\fntext[now]{Present address: \it Kavli IPMU, Kashiwa, 277-8583, Japan}
\author[address_icepp]{S.~Kobayashi}
\author[address_icepp]{N.~Matsuzawa}
\author[address_icepp]{T.~Mori}
\author[address_icepp]{S.~Ogawa}
\author[address_icepp]{R.~Onda}
\author[address_icepp]{W.~Ootani\corref{mycorrespondingauthor}}
\cortext[mycorrespondingauthor]{Corresponding author}
\ead{wataru@icepp.s.u-tokyo.ac.jp}
\author[address_icepp]{R.~Sawada}

\author[address_HPK]{K.~Sato}
\author[address_HPK]{R.~Yamada}

\address[address_icepp]
{International Center for Elementary Particle Physics (ICEPP), The University of Tokyo,
Tokyo, 113-0033 Japan}
\address[address_HPK]
{Solid State Division, Hamamatsu Photonics K. K., Hamamatsu-City, 435-8558 Japan}

\begin{abstract}
   A large-area Multi-Pixel Photon Counter (MPPC) sensitive to vacuum ultra violet (VUV) light 
   has been developed for the liquid xenon (LXe) scintillation detector of the MEG II experiment.
   The LXe detector is designed 
   to detect the 52.8\,MeV photon
   from the lepton flavour violating decay $\megsign$ and is based on $900\,\ell$ LXe with a highly granular 
   scintillation readout by 4092 VUV-MPPCs with an active area of $139\,\mathrm{mm}^2$ each, totalling $0.57\,\mathrm{m}^2$.
   The VUV-MPPC shows an excellent performance in LXe, which includes
   a high photon detection efficiency (PDE) up to 21\% 
   for the LXe scintillation light in the VUV range, a high gain,
   a low probability of the optical cross-talk and the after-pulsing, a low dark count rate and a good single photoelectron resolution.
   The large active area of the VUV-MPPC is formed by connecting 
   four independent small VUV-MPPC chips in series to avoid the increase of the sensor capacitance 
   and thus, to have a short pulse-decay-time, 
   which is crucial for high rate experiments.
   Performance tests of 4180 VUV-MPPCs produced for the LXe detector 
   were also carried out at room temperature prior to the installation to the detector
   and all of them with only a few exceptions were found to work properly.
   The design and performance of the VUV-MPPC are described in detail as well as the results 
   from the performance tests at room temperature.
\end{abstract}


\end{frontmatter}

\tableofcontents

\section{Introduction}

Liquid rare-gas scintillation detectors are employed extensively in many different fields such as particle and nuclear physics and medical sciences.
The major difficulty with the usage of liquid rare-gas as the scintillator medium 
is the detection of the scintillation light in the vacuum ultra violet (VUV) range; 
the emission peak at $\lambda=175\,\mathrm{nm}$ for liquid xenon (LXe)~\cite{Nakamura:LXeEmissionSpectrum},
$147\,\mathrm{nm}$ for liquid krypton (LKr)~\cite{Aprile:2008bga} and 
$128\,\mathrm{nm}$ for liquid argon (LAr)~\cite{Aprile:2008bga}.
VUV enhanced photomultiplier tubes (PMTs) or PMTs with wavelength shifting materials have been often
employed to detect the liquid rare-gas scintillation light. 

The silicon photomultiplier (SiPM) is a rapidly emerging photosensor technology 
as a possible replacement of the PMT.
The SiPM is now a major candidate for the photosensor of 
scintillation detectors.
The LXe scintillation detector for the MEG II experiment~\cite{MEG-design-paper}
is designed to precisely measure the energy, position and time of 
the 52.8\,MeV photon from $\megsign$, based on $900\,\ell$ LXe with a high-granularity scintillation readout by 4092 SiPMs.
A large-area Multi-Pixel Photon Counter (MPPC)
\footnote{MPPC is the product name of the SiPM produced by Hamamatsu Photonics K.K.}
with an enhanced VUV sensitivity has been
developed for the LXe detector
since the currently commercially available SiPMs are not sensitive to the VUV light. 
A prototype model of the VUV-MPPC had been successfully tested
and shown an excellent performance in LXe as previously reported in Ref.~\cite{Ootani:2015cia}.
A production model 
to be used for the LXe detector 
has been developed where the main improvement is a significant reduction of the optical cross-talk.
This article describes in detail the design of the VUV-MPPC and the performance measured in LXe mainly for the production model. 
The results from the performance tests at room temperature for the 4180 VUV-MPPCs produced for the LXe detector are also presented.

\section{Design of VUV-enhanced MPPC}
\label{sec:Design of VUV-enhanced MPPC}

Good resolutions of the MEG II LXe detector for the 52.8\,MeV photon from $\megsign$ 
can be achieved by a highly granular and uniform scintillation readout
with VUV-sensitive MPPCs.
The front side of the detector of $0.92\,\mathrm{m}^2$ on which 
the signal photon impinges is tiled with 4092 VUV-MPPCs 
with a total active area of $0.57\,\mathrm{m}^2$.
The major requirements for the VUV-MPPC of the MEG II experiment
can be summarised as
the photon detection efficiency (PDE) for the LXe scintillation light higher than 10\%, 
the gain higher than $5\times 10^5$, 
the probability of the optical cross-talk and after-pulsing lower than 15\% each,
the dark count rate lower than $10\,\mathrm{Hz}/\mathrm{mm}^2$ at LXe temperature
and the capability of single photoelectron resolution.

\begin{figure}[htpb]
   \begin{center}
      \includegraphics[width=6cm]{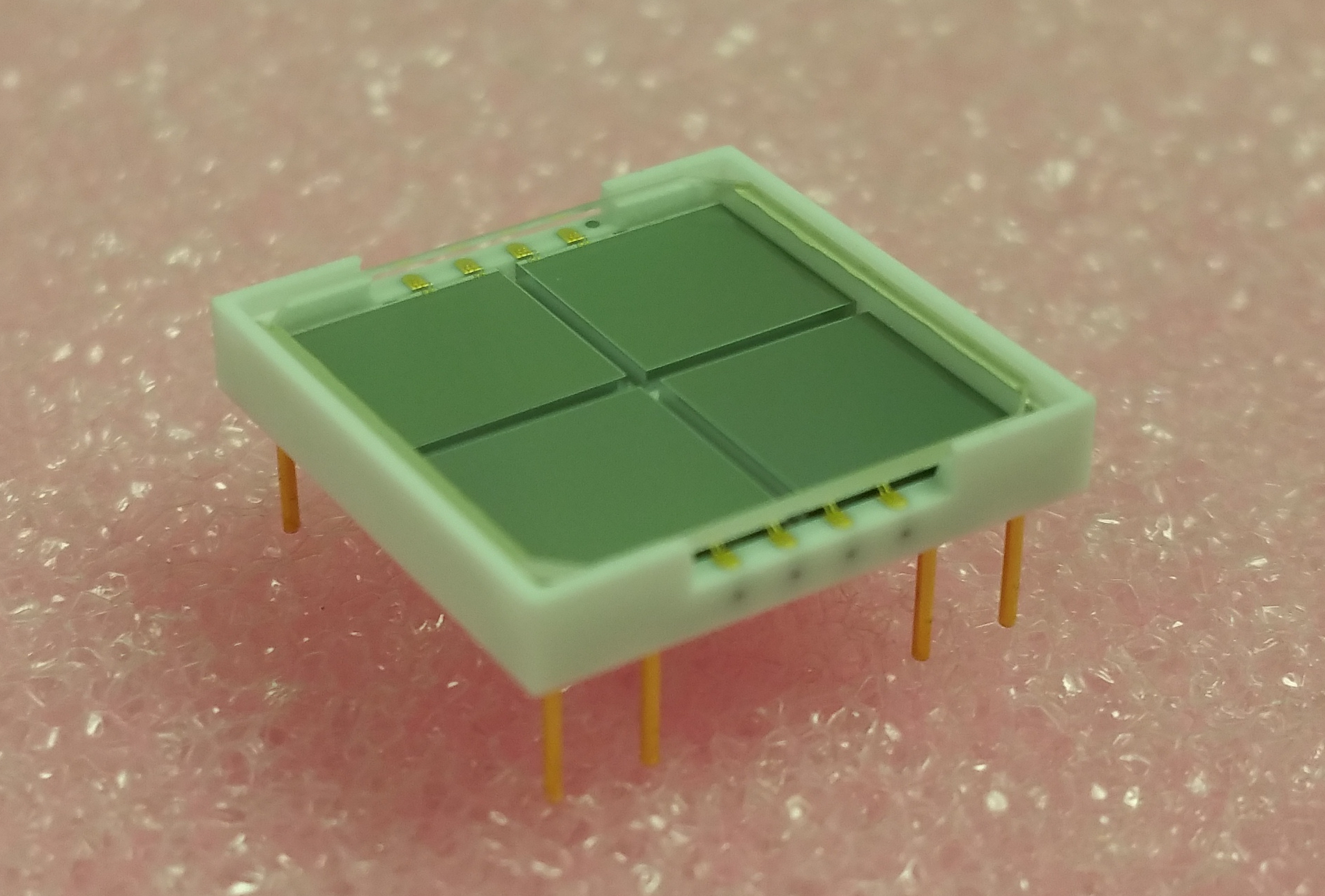}
      \includegraphics[width=7cm]{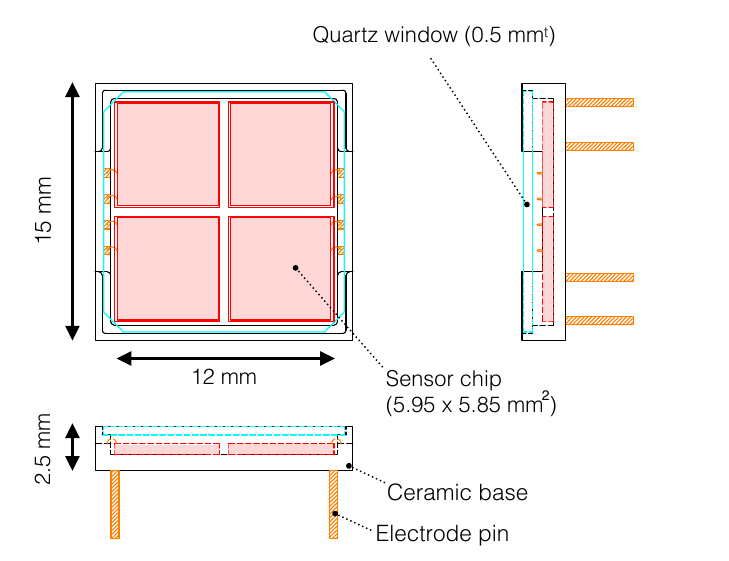}
   \end{center}
   \caption{Large-area VUV-sensitive MPPC developed for the MEG II LXe detector 
      (Hamamatsu Photonics K.K., S10943-4372). }
   \label{fig:S10943-4372}
\end{figure}

Fig.\,\ref{fig:S10943-4372} shows the photograph and 
the schematic view of the production model of the VUV-MPPC 
(Hamamatsu Photonics K.K. S10943-4372).
It is a discrete array of four bare MPPC chips, each of which has an active area 
of $5.95\times 5.85\,\mathrm{mm}^2$ with a micro-cell pitch of $50\,\mu\mathrm{m}$.
The four chips are glued on a ceramic base 
with low-temperature-resistant conductive adhesive with a spacing of $0.5\,\mathrm{mm}$
and can be readout individually.
In the MEG II LXe detector, the four chips are connected in series in a readout printed circuit board (PCB)
as described in Sec.\,\ref{sec:Performance of Large-area MPPC} to form a single sensor 
with a total active area of $139\,\mathrm{mm}^2$
\footnote{
``Chip'' and ``sensor'' shall hereafter mean the single MPPC chip 
with an active area of $5.95\times 5.85\,\mathrm{mm}^2$ 
and the sensor package composed of the four MPPC chips, respectively, 
unless otherwise noted.}.
The outline dimension of the sensor is 
$15\,\mathrm{mm}\times 15\,\mathrm{mm}\times 2.5\,\mathrm{mm}$.
The four chips in the sensor are covered by a $0.5\,\mathrm{mm}$-thick VUV-transparent quartz window for protection, 
which is a non-hermetic window with thin slits at the two sides of the package.
The thin gap of $0.5\,\mathrm{mm}$ between the window and the chips 
is filled with LXe when the sensor is operated in LXe. 
The refractive index of LXe at the emission peak
of the LXe scintillation light is $1.64$~\cite{NakamuraLXeRefractiveIndex},
which is close to that of the quartz window ($1.60$).
The reasonable matching of the refractive indices helps to minimise an 
undesirable reflection at the boundary between the quartz window and LXe.

The sensitivity to the VUV light of the commercial SiPM is quite low 
since the VUV light is absorbed in a thick protection coating on top of the SiPM chip,
usually made of epoxy resin.
Even without the protection coating, 
the VUV light is absorbed in the top contact layer of the SiPM chip
before reaching the active layer 
due to the extremely short attenuation length of the LXe scintillation light
in silicon of about 5\,nm.
The high sensitivity of the VUV-MPPC to the LXe scintillation light 
is achieved by removing the protection coating,
optimising the optical matching between the LXe and the chip surface,
and thinning down the top contact layer.
The contact layer is, however, still 
much thicker than the attenuation length of the LXe scintillation light.
The photo-generated charge carriers in the contact layer 
diffuse and a part of the charge carriers can reach the active layer 
and trigger the avalanche amplification.
There are a few other improvements for the enhanced VUV-sensitivity.
The dopant concentration in the contact layer is adjusted 
to have low but non-zero electric field, 
causing a moderate drift of the charge carriers to the active layer.
The processing of the contact layer is optimised to minimise the lattice defects 
where the charge carriers are trapped.
More than one electron-hole pairs can be generated by a single VUV photon 
since the energy of the VUV photon is twice higher 
than the average energy required to produce an electron-hole pair, 
which is estimated to be 3.65\,eV~\cite{Kowalski-Nuclear-Electronics}.
That also contributes to the high VUV-sensitivity of the VUV-MPPC.

The correlated noises of the MPPC such as the optical cross-talk and the after-pulsing 
increase the excess noise factor and thus, worsen the resolutions. 
The after-pulsing was already significantly suppressed in the prototype model 
as reported in Ref.~\cite{Ootani:2015cia}.
The optical cross-talk has also been greatly reduced 
in the production model described in this article
by introducing a trench structure 
between the adjacent cells to prevent the luminescence photons 
generated in the avalanche in the primary cell 
from firing the neighbouring cells.
The substantial suppression of both the correlated noises enables one 
to operate it 
at a higher over-voltage
\footnote{The over-voltage ($\Delta V$) is defined as the excess voltage 
   with respect to the breakdown voltage ($V_\mathrm{bd}$).}
up to 7--8\,V as compared with the prototype model, 
which improves the PDE, the gain and the temperature stability.
The VUV-MPPC employs metal-based quench resistors 
with a much smaller temperature coefficient 
than that for the polysilicon-based quench resistors 
used in the previous models of the VUV-MPPC.
The resistance of the quench resistor at LXe temperature of 165\,K
was found to be increased only by 20\% compared to that at room temperature.
The pulse-decay-time is, therefore, kept reasonably short at LXe temperature.

\section{Performance of VUV-MPPC}
\label{sec:Performance of VUV-MPPC}

The performance of the VUV-MPPC was measured in detail at LXe temperature 
for several sensors.
In addition, operational tests were done at room temperature 
for all the produced sensors before the installation to the detector.

\subsection{Setup of Performance Measurements in LXe}
\label{sec:Setup of Performance Measurement in LXe}
\begin{figure}[htpb]
   \begin{center}
      \includegraphics[width=7cm]{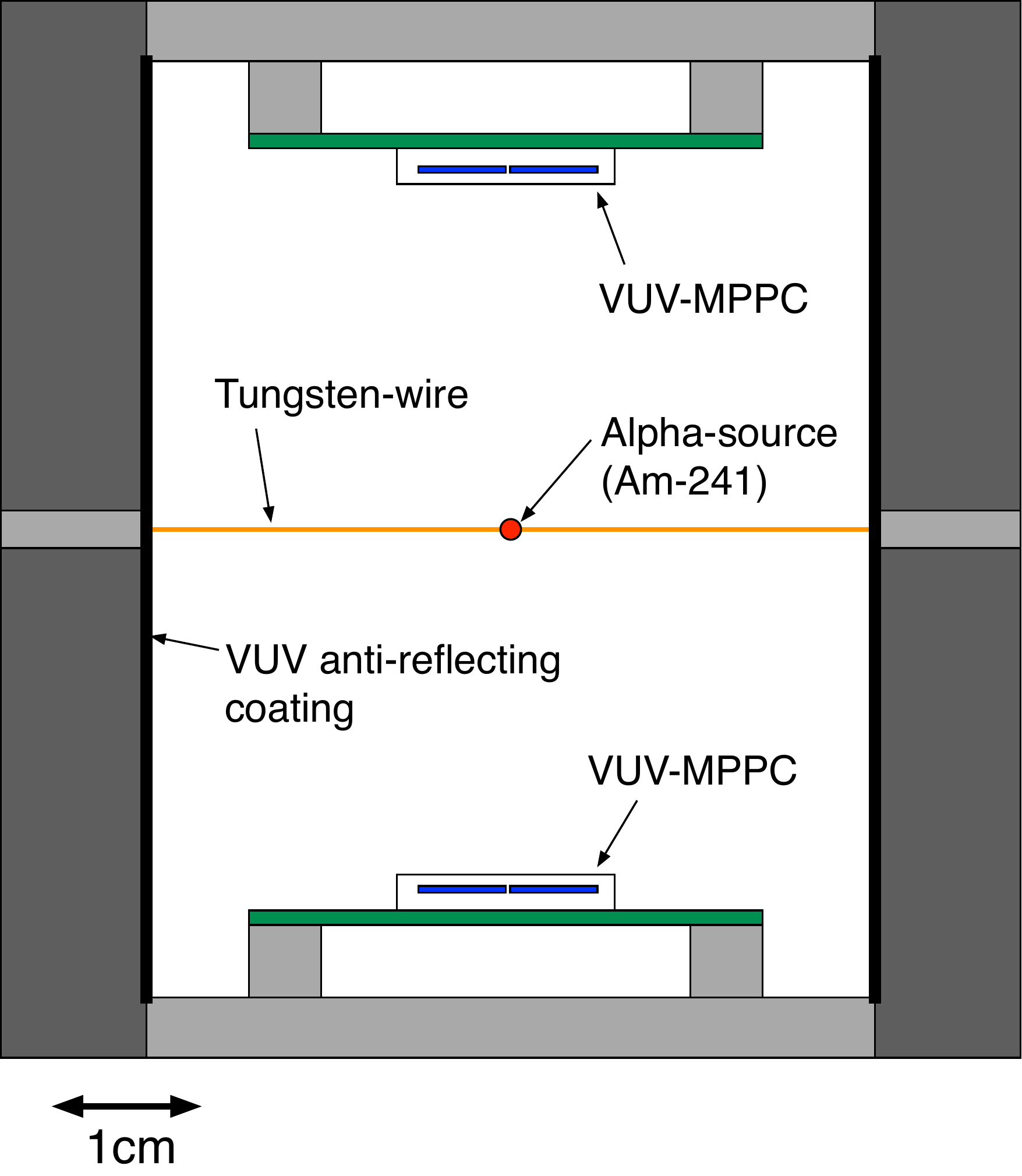}
   \end{center}
   \caption{Typical set-up for the performance measurements in LXe.}
   \label{fig:Small chamber setup}
\end{figure}

Fig.\,\ref{fig:Small chamber setup} shows a typical setup used 
for the detailed performance measurements in LXe for several sensors.
The whole setup was housed in an inner vessel of a low temperature cryostat 
where $2\,\ell$-LXe is liquefied with a pulse tube refrigerator (Iwatani PDC08).
A few sensors were mounted on each top and bottom stage 
facing a spot $\alpha$-source on a thin tungsten wire stretched in between. 
The spot $\alpha$-source was developed 
for the MEG experiment~\cite{Baldini:2006}.
A $1.5\,\mu\mathrm{m}$-thick gold foil,
in which a radioactive matrix of Am-241 in a chemical form of Americium Oxide 
with an activity of about 100\,Bq is homogeneously incorporated,
is wound around a thin gold-plated tungsten wire 
with a diameter of $100\,\mu\mathrm{m}$ 
and is covered by a 1.5\,$\mu\mathrm{m}$-thick gold foil.
The linear dimension of the spot source is about $500\,\mu\mathrm{m}$.
The energy of the $\alpha$-rays emitted from the spot source was measured in vacuum
with a silicon surface barrier detector in a different setup.
The measured energy is 4.8\,MeV with a spread of 4.0\% (in sigma)
\footnote{The measured energy spectrum has a low energy tail and the spread was evaluated 
at the right side of the peak.}, 
which is lower than the initial energy of $\alpha$-rays emitted from Am-241 of 5.5\,MeV
due to the energy loss in the gold foil surrounding the source.
The sensors on the top and bottom stages were illuminated 
by the LXe scintillation light generated by the spot $\alpha$-source, 
which can be considered as a point light source
because of the range of $\alpha$-ray in LXe as short as $40\,\mu\mathrm{m}$.
A few blue LEDs (OSA Opto Light, OCU-400 UE390) 
were also installed in the setup for a calibration purpose.
The LXe volume was surrounded by a cylindrical wall 
with a VUV anti-reflecting coating (Acktar black~\cite{AcktarBlack}) 
to minimise the influence of the reflection 
of the scintillation light from the $\alpha$-source
in the PDE measurement.
A reliable prediction of the number of scintillation photons 
impinging on the sensors is required in the PDE measurement as described later.
The signal from the sensor was transmitted through a vacuum feedthrough 
to an amplifier placed outside the cryostat and then to a waveform digitizer (DRS4~\cite{Ritt:2010zz}).

\subsection{Performance of Single Chip in LXe}
\label{sec:Performance of Single MPPC Chip}

\begin{figure}[htpb]
   \begin{center}
      \includegraphics[width=8cm]{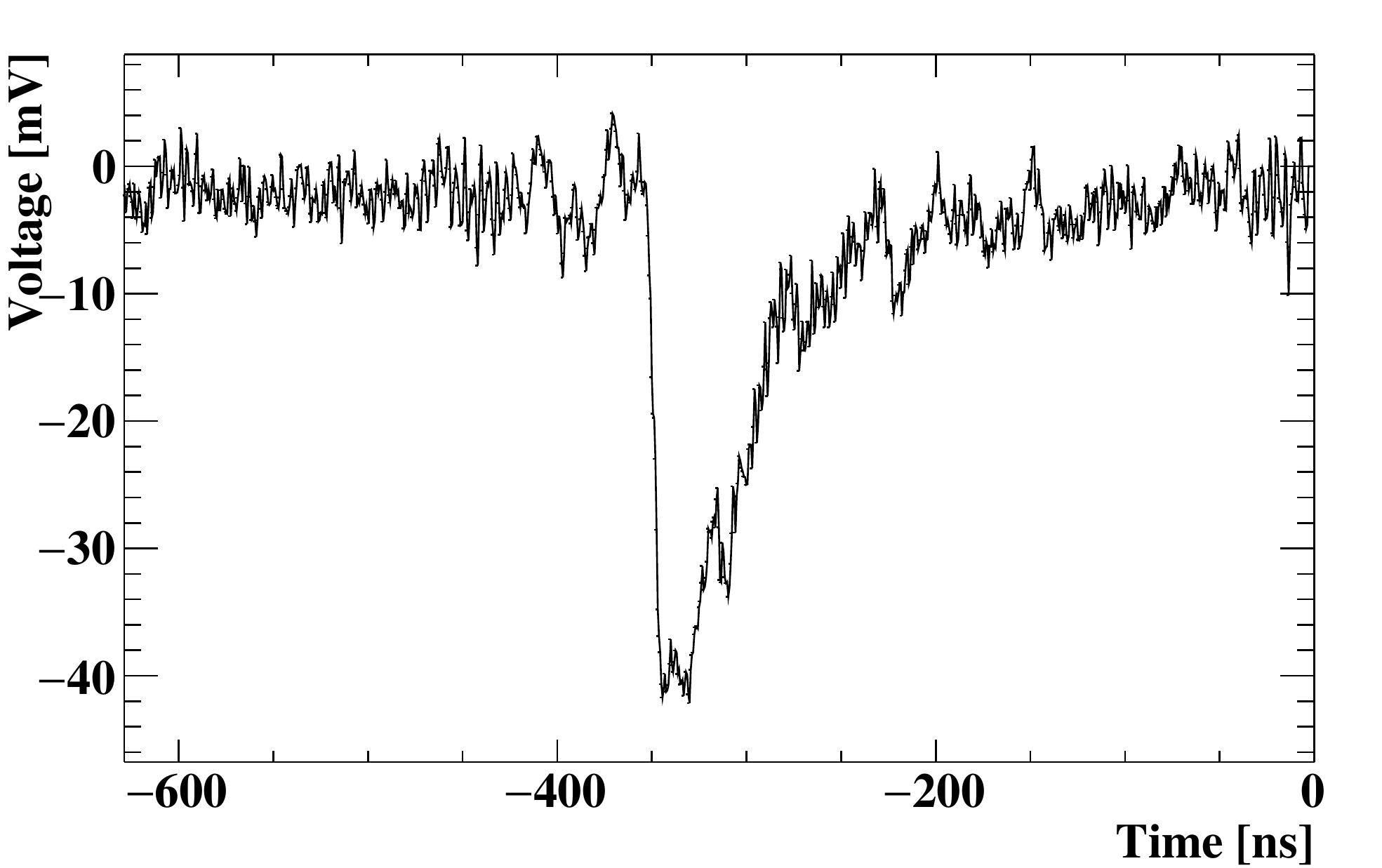}
   \end{center}
   \caption{Typical single photoelectron signal from a single chip 
      of $5.95\times 5.85\,\mathrm{mm}^2$ operated in LXe.}
   \label{fig:MPPC signal 6x6}
\end{figure}

The detailed properties of the VUV-MPPC were measured in LXe
for individual chips of $5.95\times 5.85\,\mathrm{mm}^2$.
Fig.\,\ref{fig:MPPC signal 6x6} shows a typical waveform 
for the single photoelectron signal. 
The rise and decay times of the single photoelectron signal 
were measured by fitting exponential functions to be $2.8\,\mathrm{ns}$ and $86\,\mathrm{ns}$, respectively.

\begin{figure}[htpb]
   \begin{center}
      \includegraphics[width=8cm]{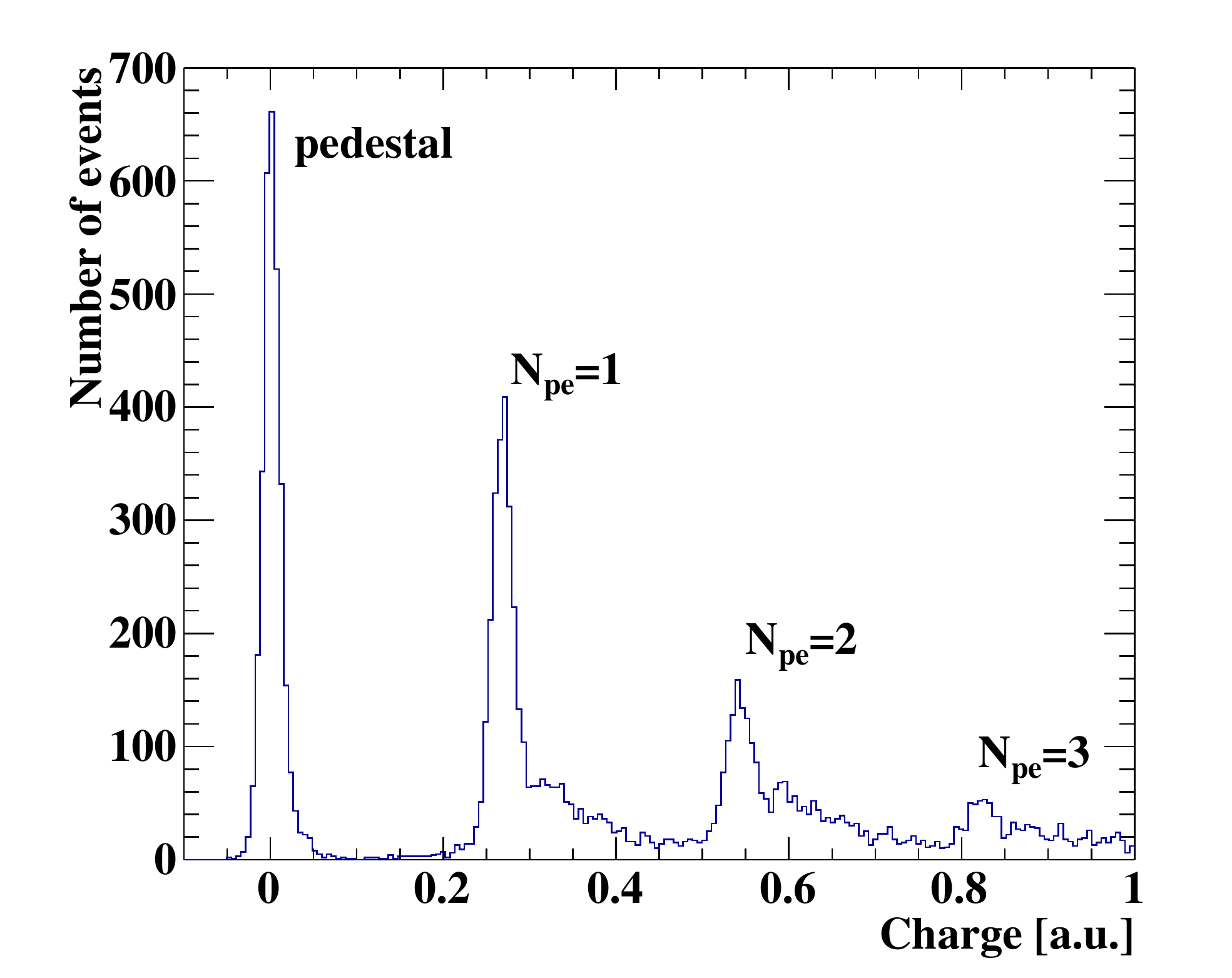}
   \end{center}
   \caption{
  Charge distribution obtained for low-level LED light pulses 
  with the single MPPC chip of $5.95\times 5.85\,\mathrm{mm}^2$ 
  operated in LXe at $\Delta V=7\,\mathrm{V}$.
   }
   \label{fig:MPPC single photoelectron distribution 6x6}
\end{figure}

The charge distribution obtained with a single chip operated 
at $\Delta V=7\,\mathrm{V}$ for low-level light pulses
from the blue LED is shown 
in Fig.\,\ref{fig:MPPC single photoelectron distribution 6x6}.
The peaks for successive numbers of photoelectrons are clearly resolved.
The tail at the right side of each photoelectron peak comes from the pileup due to after-pulsing.

\begin{figure}[htpb]
   \begin{center}
      \includegraphics[width=8cm]{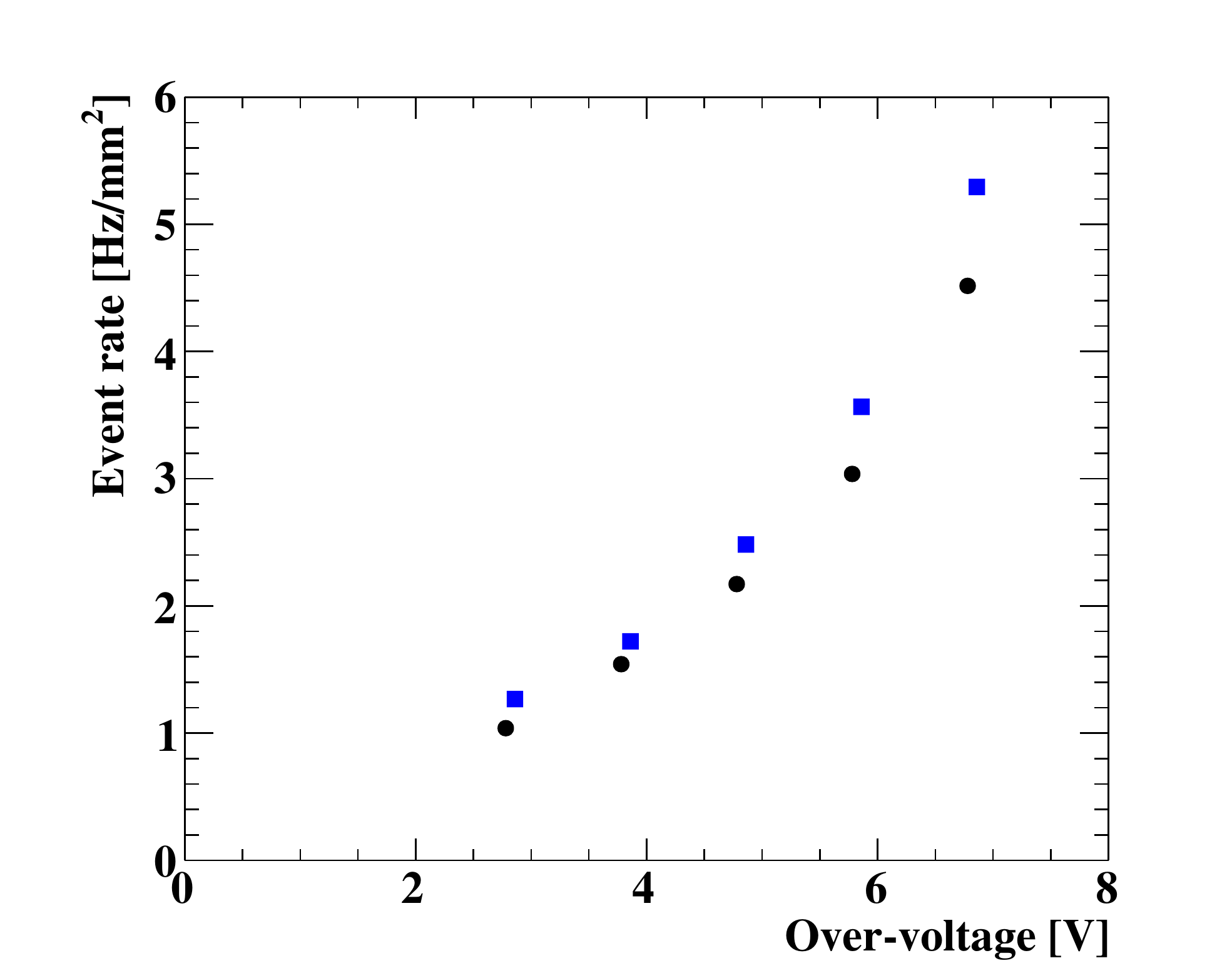}
   \end{center}
   \caption{
  Measured dark count rate as a function of the over-voltage 
  for two different single MPPC chips operated in LXe. 
   }
   \label{fig:DCR vs OV}
\end{figure}

\begin{figure}[htpb]
   \begin{center}
      \includegraphics[width=8cm]{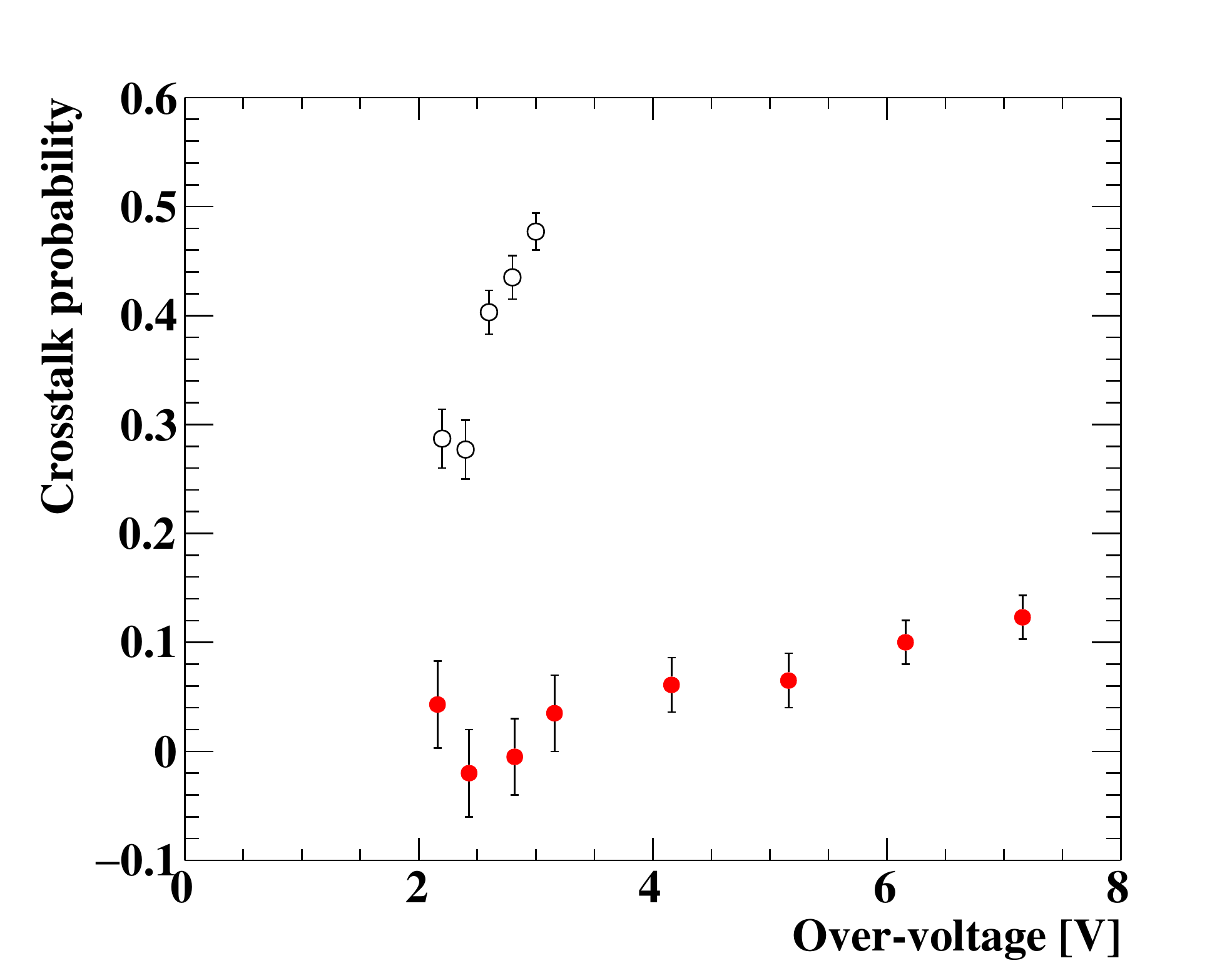}
   \end{center}
   \caption{
  Measured probability of the optical cross-talk
  as a function of the over-voltage (red filled circle). 
  The probability for the prototype model 
  is also shown for comparison (black open circle).
}
   \label{fig:cross-talk}
\end{figure}

\begin{figure}[htpb]
   \begin{center}
      \includegraphics[width=8cm]{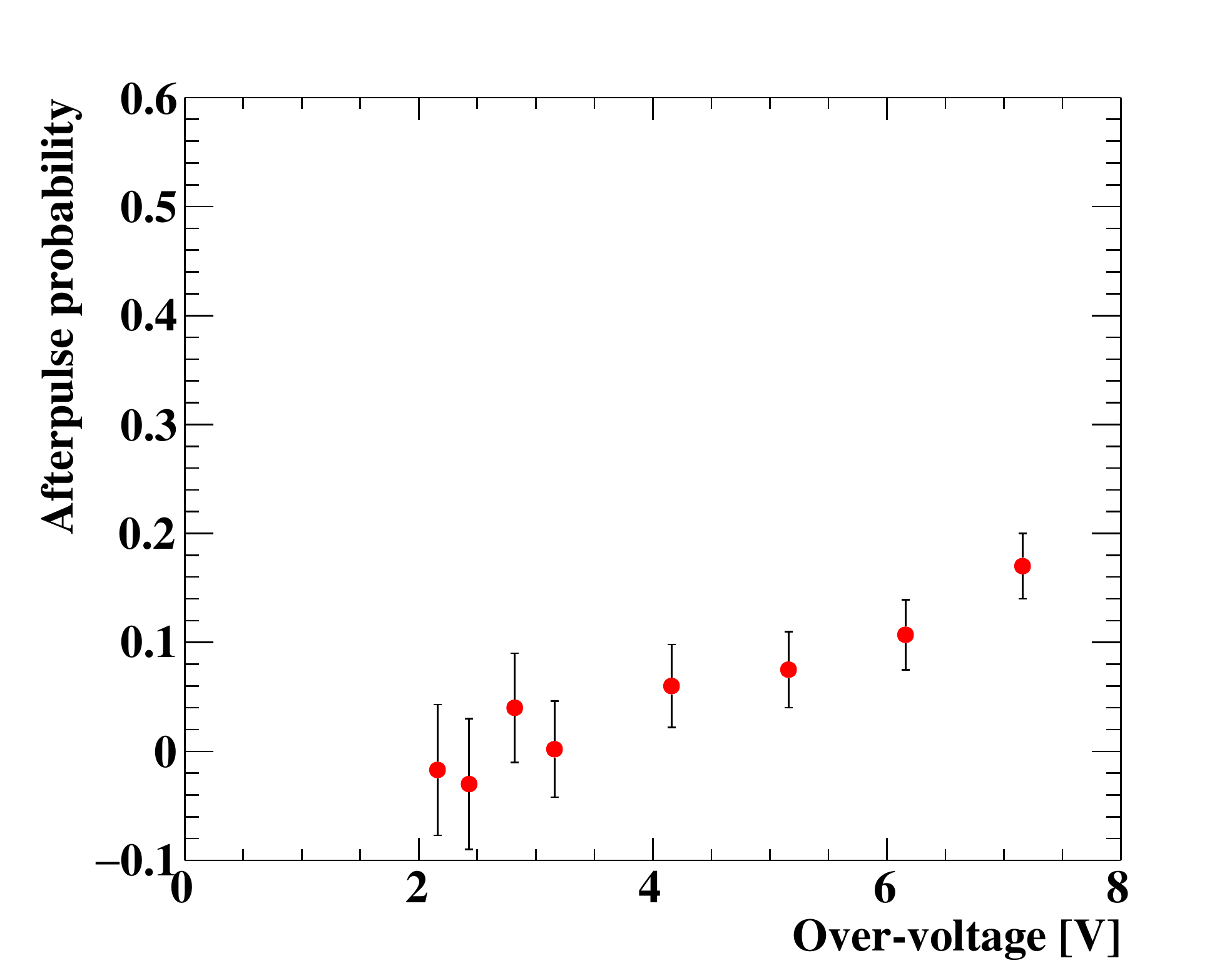}
   \end{center}
   \caption{
  Measured probability of the after-pulsing
  as a function of the over-voltage. 
   }
   \label{fig:after-pulsing}
\end{figure}

The dark count rate as a function of the over-voltage was measured at LXe temperature 
by counting the random single photoelectron events
as shown in Fig.\,\ref{fig:DCR vs OV}.
The scintillation light events caused by environmental radio-activity and cosmic-rays 
were vetoed by eliminating coincident events over multiple chips.
Some of the scintillation light events with low light level, however, were left unvetoed 
because of the non-negligible probability of zero photoelectron events.
Furthermore, the measured rate includes the single photoelectron events due to the after-pulsing.
The typical dark count rate was estimated to be $5\,\mathrm{Hz}/\mathrm{mm}^2$ at $\Delta V = 7\,\mathrm{V}$, 
although it was somewhat overestimated for the reasons mentioned above. 
This extremely low dark count rate, which is about five orders of magnitude lower than the rate at room temperature,
enables one to operate the VUV-MPPC 
with a good single photoelectron resolution 
even with a large area of $139\,\mathrm{mm}^2$ as described in Sec.\,\ref{sec:Performance of Large-area MPPC}.
The probabilities of the correlated noises 
were evaluated with low-level LED light pulses
by means of the method described in Ref.~\cite{T2K:NIMA2009}.
Firstly, the probability of the true one photoelectron event 
without the correlated noises is estimated from the probability 
of the zero photoelectron event assuming the Poisson distribution.
The observed probability of the one photoelectron event 
is lower than estimated from the zero photoelectron probability 
because some of the true one photoelectron events are observed 
as the events with two or more photoelectrons
due to the overlap of the correlated noises.
The probabilities of the correlated noises 
are then calculated from the difference 
between the observed and estimated probabilities of the one photoelectron event.
The overall probability of the optical cross-talk and the after-pulsing 
is estimated with a charge integration time of 150\,ns 
which is long enough to contain most of the after-pulsing with a delayed timing, 
while the probability of the optical cross-talk is separately estimated 
with a shorter integration time of 30\,ns.
Fig.\,\ref{fig:cross-talk} and Fig.\,\ref{fig:after-pulsing} 
show the measured probabilities 
of the optical cross-talk and the after-pulsing 
as a function of the over-voltage, respectively.
In Fig.\ref{fig:cross-talk} the cross-talk measured for the prototype model is also shown for comparison.
The cross-talk is drastically reduced for the production model, 
which enables one to operate it at a much higher over-voltage up to 7--8\,V,
where the overall probability of the correlated noises is still as low as $30$\%.

\begin{figure}[htpb]
   \begin{center}
      \includegraphics[width=8cm]{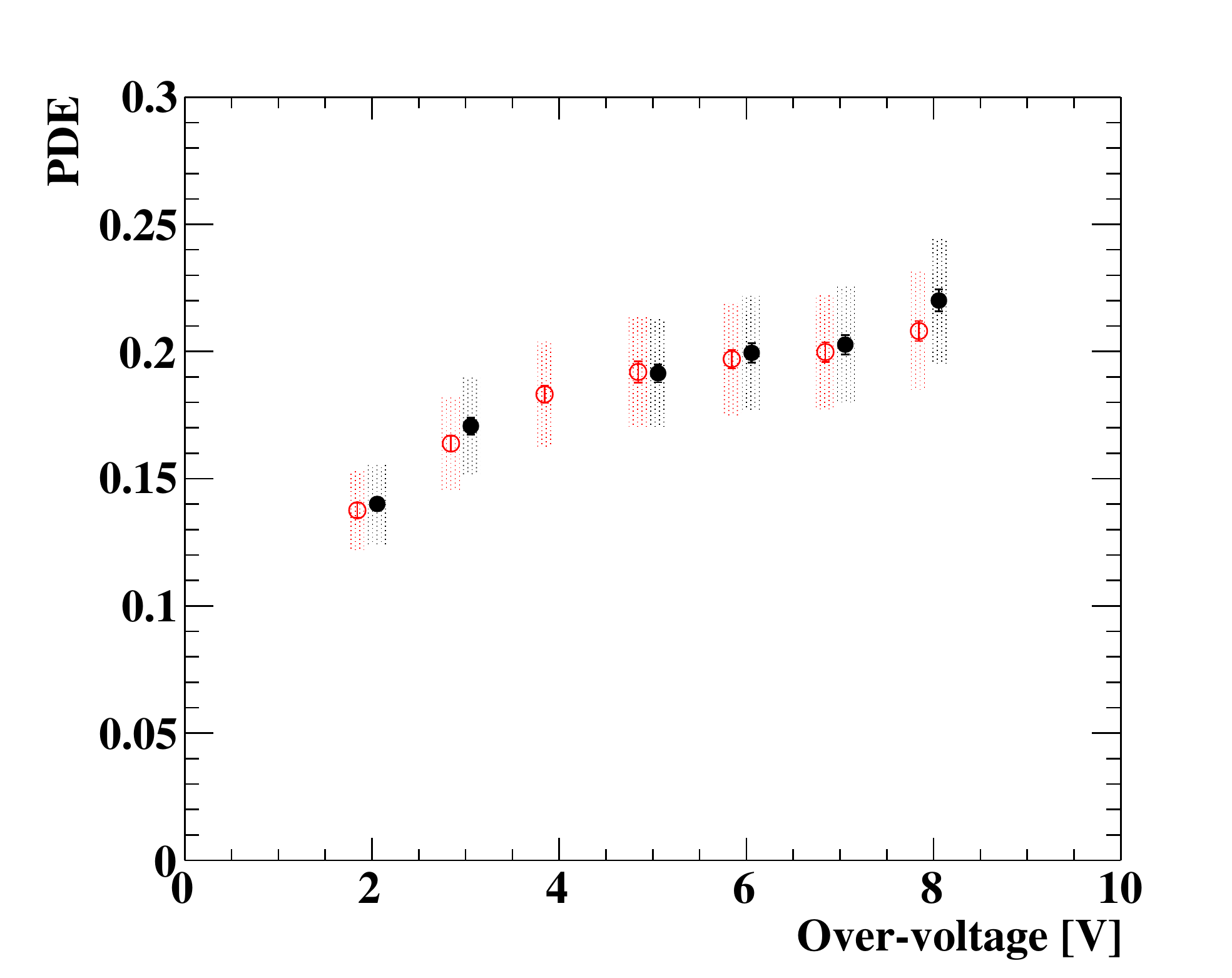}
   \end{center}
   \caption{
  Measured PDE for two VUV-MPPC chips as a function of the over-voltage.
  The error bars and the strips represent the statistical and the systematic errors, respectively.
  The contributions from the correlated noises are corrected for using the measured probabilities.
   }
   \label{fig:PDE}
\end{figure}

The PDE of the VUV-MPPC was measured
as the ratio of the observed number of photoelectrons 
to the expected number of photons impinging on the active area 
of the MPPC chip.
The contributions from the correlated noises are subtracted from the observed number of photoelectrons 
using the measured probabilities.
The expected number of photons was evaluated from
the solid angle subtended by the active area of the chip to the spot $\alpha$-source
assuming the scintillation light yield of the $\alpha$-ray of 
$(51\pm5)\times10^3\,\mathrm{photons}/\mathrm{MeV}$~\cite{Doke:1999ku}.
The contribution from the reflected light on the surrounding wall 
with the VUV anti-reflecting coating was evaluated to be negligibly small,
while the reflection on the surface of the gold foil 
surrounding the spot $\alpha$-source was found to be non-negligible 
and taken into account in the estimation of the PDE.
Fig.\,\ref{fig:PDE} shows the measured PDE 
for two chips as a function of the over-voltage
where the systematic error of the measurement comes mainly from
the error in the scintillation light yield 
for $\alpha$-rays (10\%~\cite{Doke:1999ku})
and the uncertainty in the estimation of the effect of the reflection 
on the surface of the spot $\alpha$-source (5\%).
The PDE was measured to be 14--21\% depending on the over-voltage,
which sufficiently fulfils 
the requirement for the MEG II LXe detector.

\subsection{Performance of Large-area Sensor in LXe}
\label{sec:Performance of Large-area MPPC}
\begin{figure}[htpb]
   \begin{center}
      \includegraphics[width=7cm]{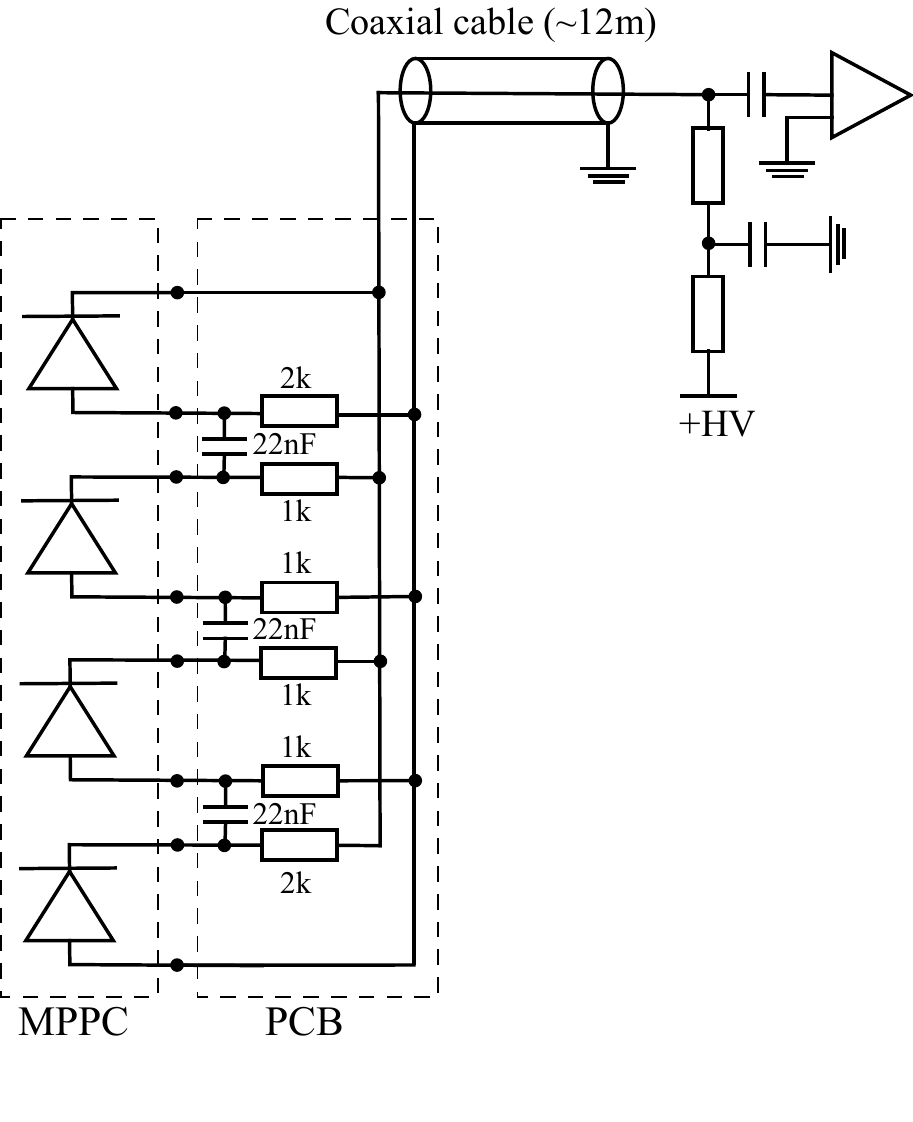}
   \end{center}
   \caption{
  Scheme of the series connection of the four MPPC chips on the VUV-MPPC.
   }
   \label{fig:series connection}
\end{figure}

While a highly granular scintillation readout is required for the MEG II LXe detector, 
the maximum size of the active area of the standard MPPC 
($6\times 6\,\mathrm{mm}^2$)
is even too small to manage the corresponding enormous number of readout channels.
The optimal sensor size for the LXe detector is found to be about $10\times 10\,\mathrm{mm}^2$.
The large sensor capacitance of 5\,nF expected for that sensor size will be an issue due to 
the resultant long pulse-decay-time.
The signal readout by transimpedance amplifiers with a low input-impedance to mitigate the long pulse-decay-time
is not possible for the LXe detector because of the non-negligible impedance of the signal cable as long as 12\,m 
between the sensor and the amplifier.
The long cable is required to transmit the signal 
from the inside of the detector cryostat to the readout electronics placed outside the cryostat.
A single sensor with a large active area of $139\,\mathrm{mm}^2$ is, therefore, formed 
by connecting four independent small MPPC chips ($5.95\times 5.85\,\mathrm{mm}^2$ each) in series
to avoid the increase in the sensor capacitance. 
The scheme of the series connection, which is  made on a signal readout PCB underneath the sensor, 
is illustrated in Fig.\,\ref{fig:series connection}.
The four chips are connected in series with decoupling capacitors in between 
such that the four chips are biased by a common voltage.
This connection scheme is more advantageous than the simple series connection.
The bias voltage is the same as the one for the single chip 
in contrast to the four times higher bias voltage required 
for the simple series connection.
There is no voltage difference between the adjacent chips, 
which eliminates the risk of the discharge between the adjacent chips.
Four chips with almost the same breakdown voltage are selected 
to assemble each sensor in the production by the company
such that the four chips within each sensor have the same gain when biased by a common voltage.

\begin{figure}[htpb]
   \begin{center}
      \includegraphics[width=8cm]{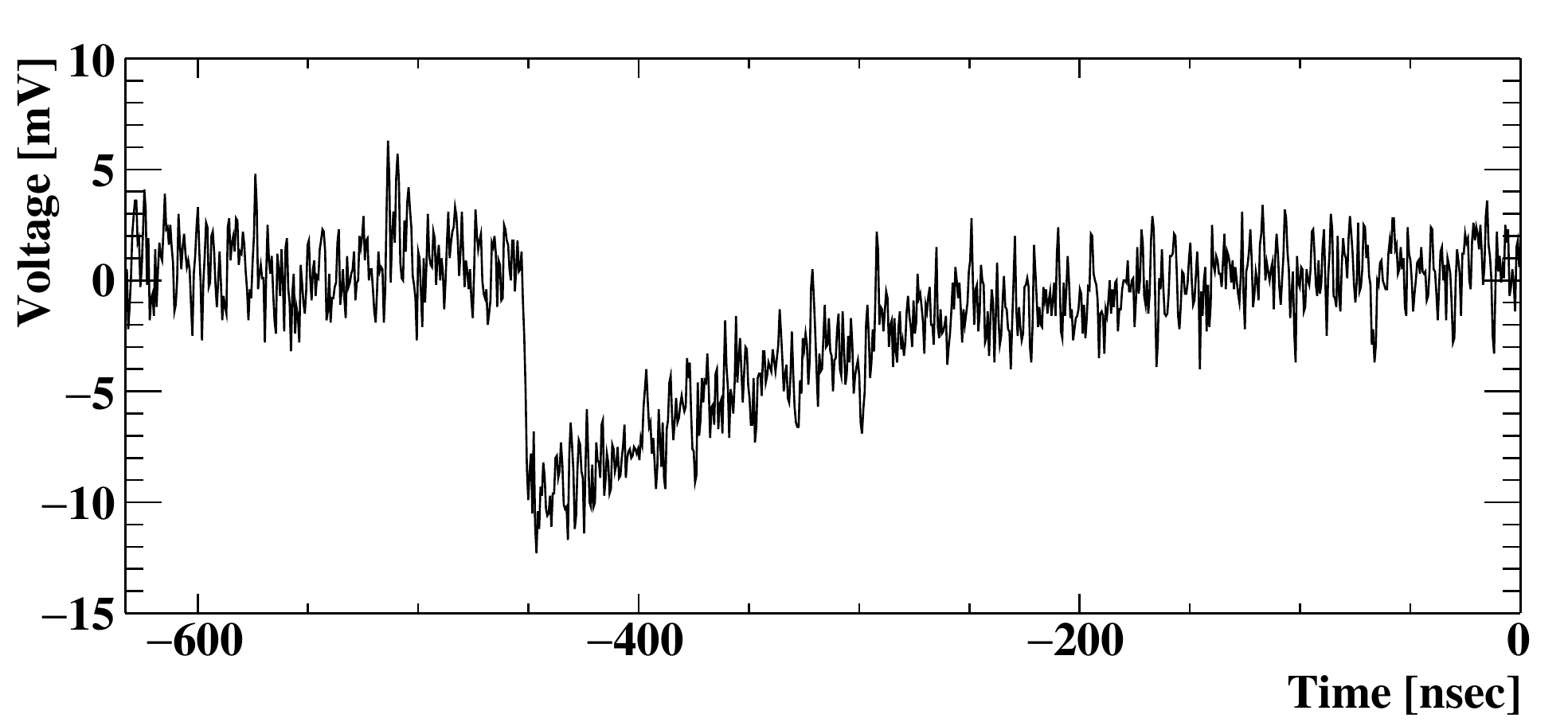}
      \includegraphics[width=8cm]{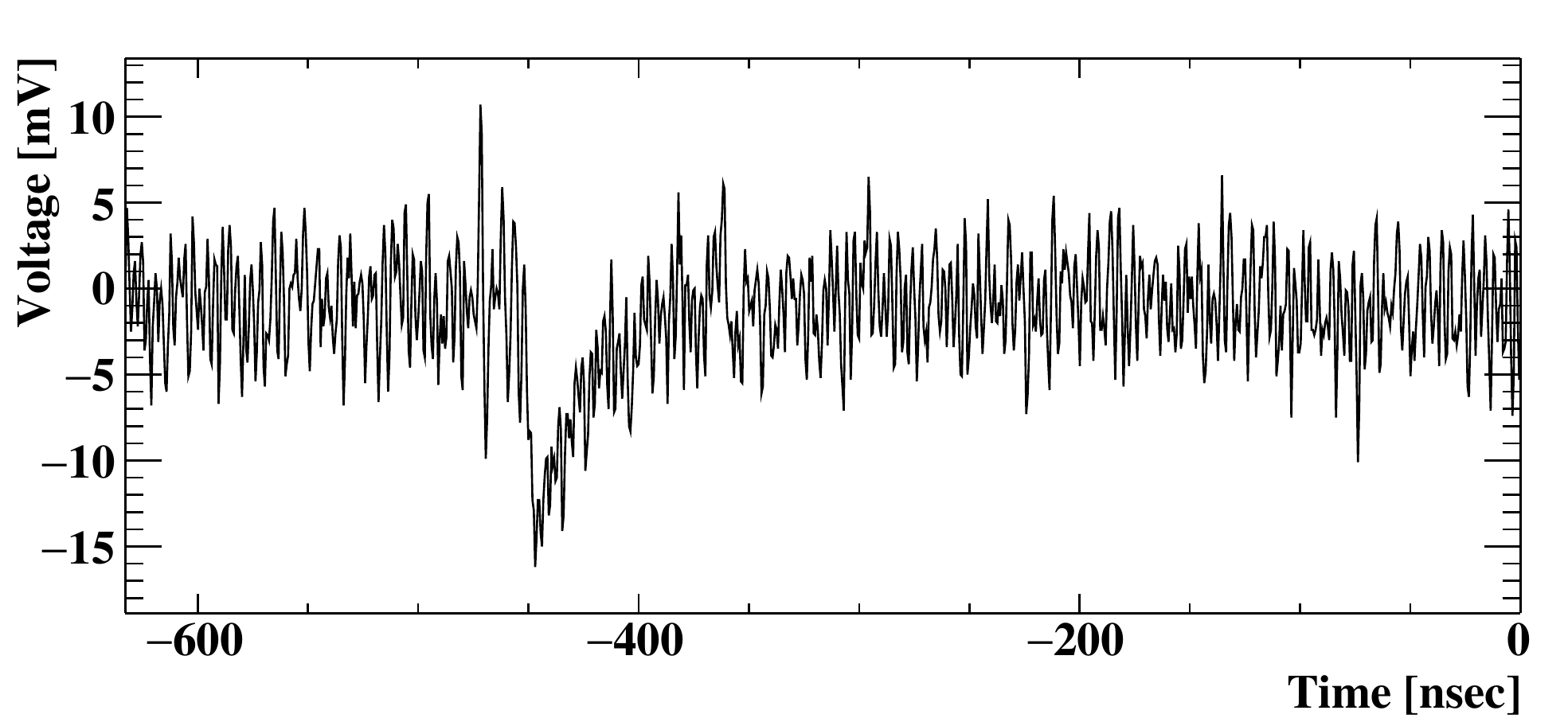}
   \end{center}
   \caption{
  Typical waveforms for the signal with two photoelectrons
  (primary photoelectron overlapped with optical cross-talk) 
  from the VUV-MPPC with the four chips connected
  in parallel (top) and in series (bottom).
  The VUV-MPPCs were operated in LXe. 
   }
   \label{fig:MPPC signal 12x12}
\end{figure}

\begin{figure}[htpb]
   \begin{center}
      \includegraphics[width=8cm]{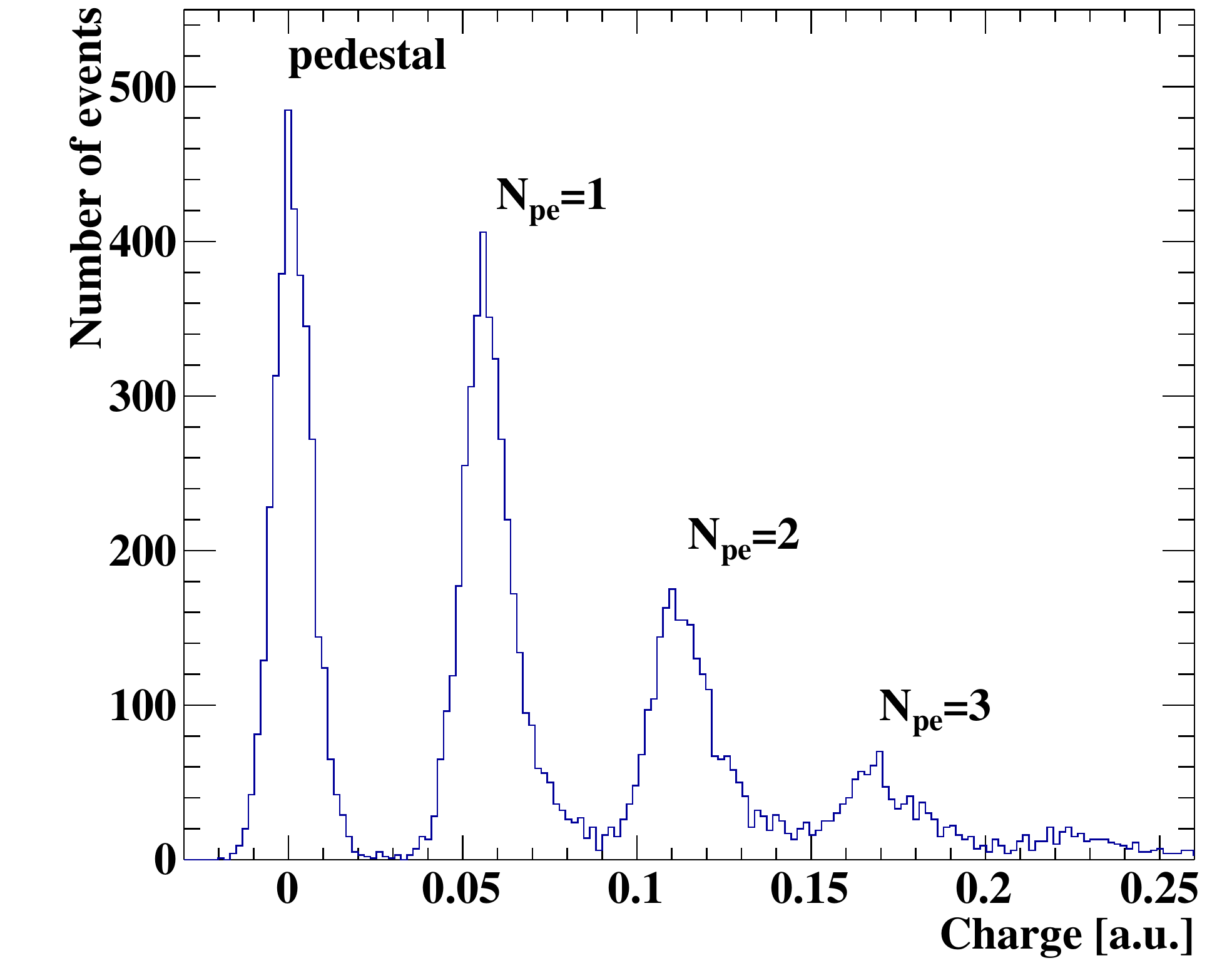}
   \end{center}
   \caption{
  Charge distribution for a low-level LED light 
  for the VUV-MPPC with the four chips connected in series,
  operated in LXe at $\Delta V=7\,\mathrm{V}$. 
   }
   \label{fig:MPPC single photoelectron distribution 12x12}
\end{figure}

\begin{figure}[htpb]
   \begin{center}
      \includegraphics[width=8cm]{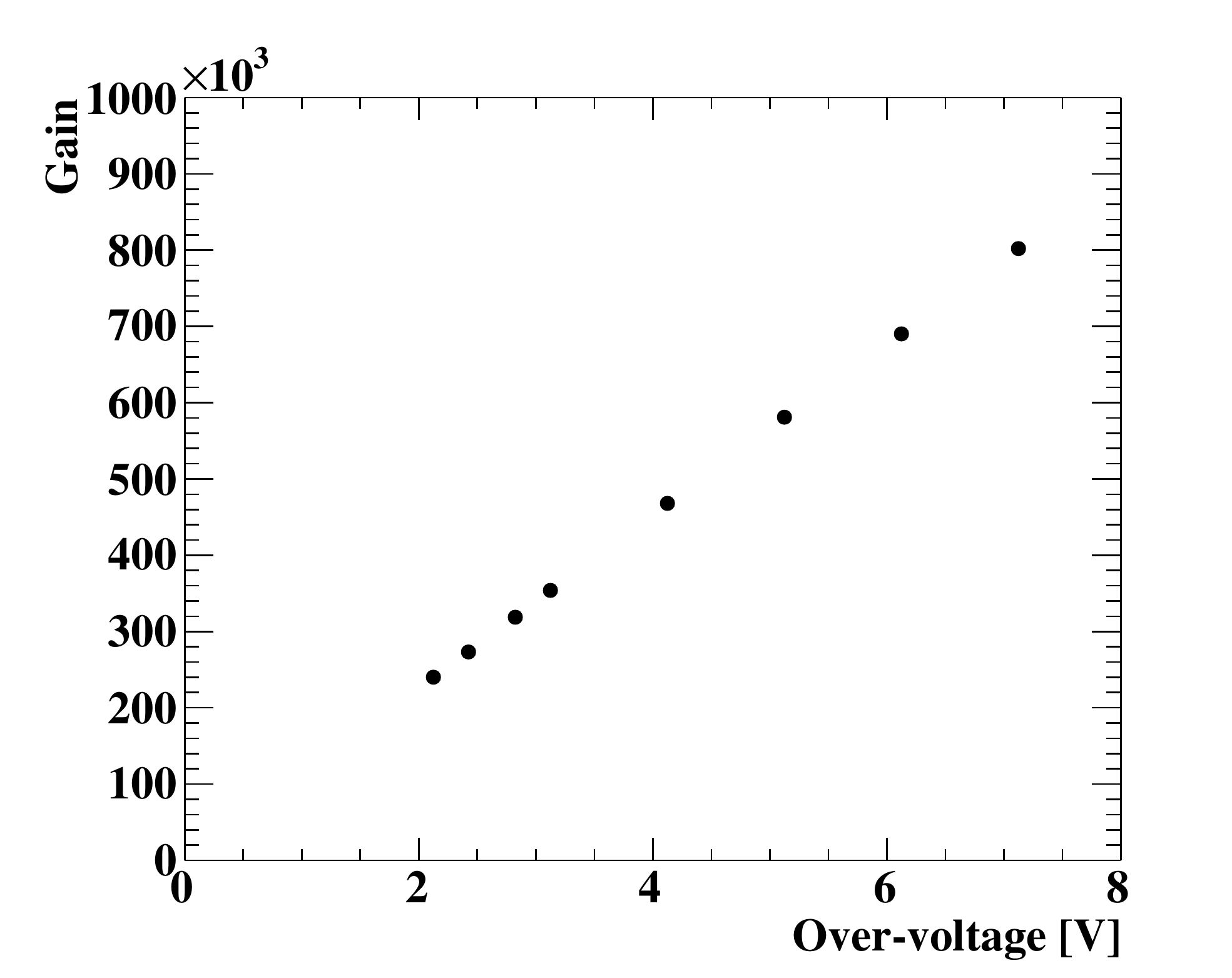}
   \end{center}
   \caption{
  Gain for the VUV-MPPC with the four chips connected in series, 
  operated in LXe as a function of the over-voltage.
   }
   \label{fig:MPPC gain 12x12}
\end{figure}

The typical waveform for the signal with two photoelectrons 
(primary photoelectron overlapped with optical cross-talk) 
from the four chips connected in series is shown 
in the bottom of Fig.\,\ref{fig:MPPC signal 12x12}.
The waveform from the four chips connected in parallel, which is equivalent 
to a single large sensor of $139\,\mathrm{mm}^2$, is also shown for comparison 
in the top of Fig.\,\ref{fig:MPPC signal 12x12}.
The pulse-decay-time for the series connection was measured 
to be $25\,\mathrm{ns}$,
which is five times shorter than that for the parallel connection.
Fig.\,\ref{fig:MPPC single photoelectron distribution 12x12} 
shows the charge distribution obtained for low-level light pulses from the blue LED 
for the four chips connected in series operated in LXe at $\Delta V=7\,\mathrm{V}$.
The photoelectron peaks are clearly resolved 
with a single photoelectron resolution as good as 12\%.
Fig.\,\ref{fig:MPPC gain 12x12} shows 
the gain computed from the single photoelectron charge measured 
as a function of the over-voltage. 
%

\begin{figure}[htpb]
   \begin{center}
      \includegraphics[width=8cm]{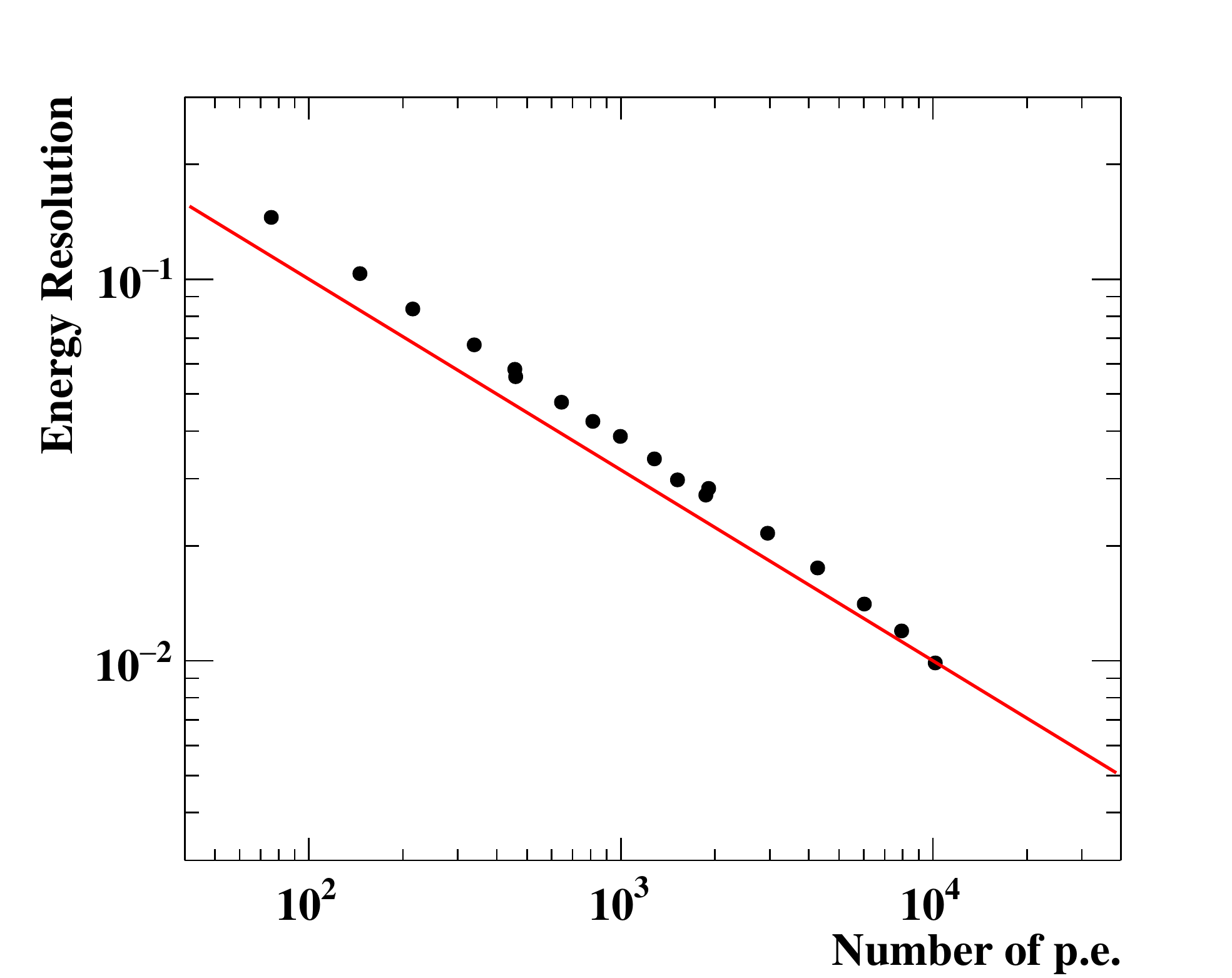}
   \end{center}
   \caption{
  Measured energy resolutions for visible light pulses from a blue-LED as a function of the detected number of photoelectrons. 
 The contribution expected from the Poisson statistics ($1/\sqrt{N_{\mathrm{p.e.}}}$) is also shown as a red solid line for comparison.
   }
   \label{fig:energy resolution visible light}
\end{figure}

The energy resolutions of the large-area VUV-MPPC were measured 
for visible light and LXe scintillation light.
The sensor was illuminated by light pulses from a blue-LED with a constant intensity 
and the resolution was measured as the standard deviation 
of the Gaussian function fitted to the charge distribution.
The measurement was repeated for different light intensities 
by varying the voltage to the LED to study 
the dependence of the resolutions on the number of photoelectrons.
The sensor was operated at a low over-voltage of 1.5\,V 
to eliminate the effect of the worsening 
of the resolution due to the correlated noises.
Fig.\,\ref{fig:energy resolution visible light} shows the measured resolutions 
as a function of the detected number of photoelectrons ($N_\mathrm{p.e.}$) 
where the contribution expected from the Poisson statistics ($1/\sqrt{N_{\mathrm{p.e.}}}$) is also shown for comparison.
The measured resolution approaches the Poisson statistics with a large number of photoelectrons. 
The deviation from the Poisson statistics with a smaller number of photoelectrons 
is considered to be due to the instability of the light intensity of the LED.

\begin{figure}[htpb]
   \begin{center}
      \includegraphics[width=8cm]{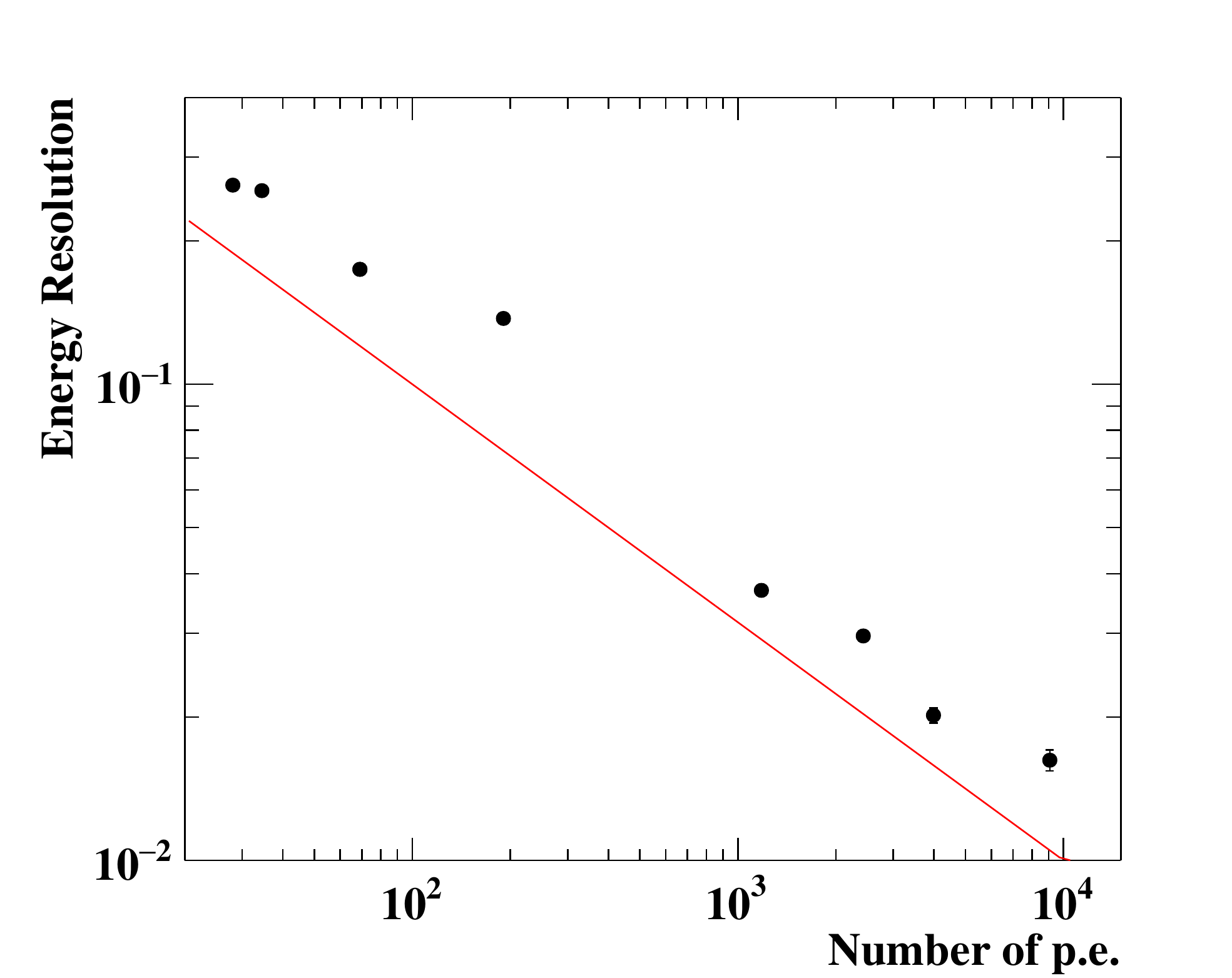}
   \end{center}
   \caption{
  Measured energy resolution for the LXe scintillation light 
  as a function of the total number of photoelectrons observed 
  by the two or four sensors used in the estimation of the resolution.
 The contribution expected from the Poisson statistics ($1/\sqrt{N_{\mathrm{p.e.}}}$) is also shown as a red solid line for comparison.
   }
   \label{fig:energy resolution VUV}
\end{figure}

The energy resolution for the LXe scintillation light was measured 
using the spot $\alpha$-source.
The prototype model of the VUV-MPPC was used in the measurements.
As described in Sec.\,\ref{sec:Performance of VUV-MPPC}, 
the spread of the energy of $\alpha$-ray emitted from the spot source
is too large to estimate the energy resolution from the spread of the charge spectrum for $\alpha$-ray events.
The resolution was, therefore, evaluated from the distribution 
of the difference in the numbers of photoelectrons observed 
by two groups, each of which consists 
of one or two sensors placed at the same distance 
from the $\alpha$-source to subtract the effect 
of the energy spread of the emitted $\alpha$-rays.
The relative resolution is estimated 
as $\sigma_{N_1-N_2}/PV_{N_1+N_2}$ and $\sigma_{N_1+N_2-N_3-N_4}/PV_{N_1+N_2+N_3+N_4}$
for the cases of two and four sensors, respectively,
where $N_i$ is the number of photoelectrons observed by the $i$-th sensor for the same $\alpha$-ray event, 
$\sigma_N$ is the standard deviation of the Gaussian function fitted to the distribution of $N$ and 
$PV_N$ is the peak value of the distribution of $N$.
The measurements were performed with several sensors placed 
at different distances from the $\alpha$-source
to study the dependence of the resolutions on the number of photoelectrons.
The sensors were operated at a low over-voltage of 1.5\,V 
to minimise the effect of the correlated noise.
Fig.\,\ref{fig:energy resolution VUV} shows the measured energy resolutions 
as a function of the total number of photoelectrons observed 
by the two or four sensors used in the estimation of the resolution.
The energy resolution for the LXe scintillation light was found to be worse
than the Poisson statistics similarly to that for the visible light 
but the deviation from the Poisson statistics is larger than that for the visible light.
The larger deviation is not understood yet, 
but could be related to the special detection process for the VUV light 
as described in Sec.\,\ref{sec:Design of VUV-enhanced MPPC}
which is quite different from that for the visible light.

\begin{figure}[htpb]
   \begin{center}
      \includegraphics[width=8cm]{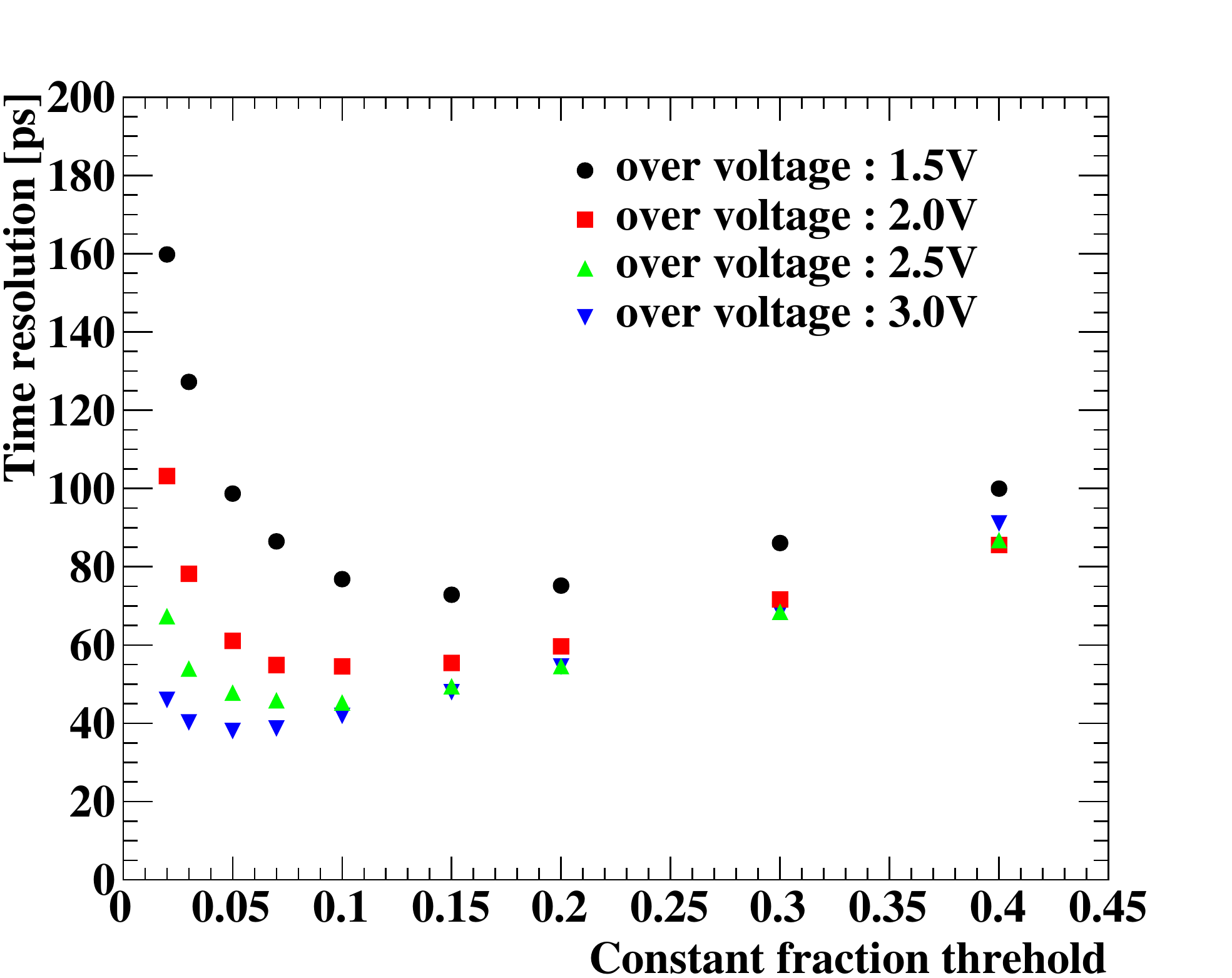}
   \end{center}
   \caption{
  Measured time resolutions for the LXe scintillation light as a function of the fraction threshold in the constant fraction method.
   }
   \label{fig:timing resolution}
\end{figure}

The time resolution for the LXe scintillation light was estimated 
from the time difference between the two sensors placed at the same 
distance from the $\alpha$ source 
where about 2000 photoelectrons were observed on average.  
The time of the MPPC signal is computed by a waveform analysis 
with the constant fraction method to eliminate the time walk effect,
where the signal time is defined as the leading edge reaches a fixed fraction of the pulse height.
Fig.\,\ref{fig:timing resolution} shows 
the measured time resolutions for different over-voltages 
as a function of the fraction threshold in the constant fraction method.
The time resolution improves for higher over voltages because of the higher signal-to-noise ratio and the higher PDE.
For each over-voltage, the best time resolution is obtained 
at a different fraction threshold.
The best overall time resolution of about 40\,ps 
is obtained at $\Delta V = 3\,\mathrm{V}$ with a fraction threshold of 5\% 
and is good enough for the MEG II LXe detector.

\begin{figure}[htpb]
    \begin{center}
      \includegraphics[width=8cm]{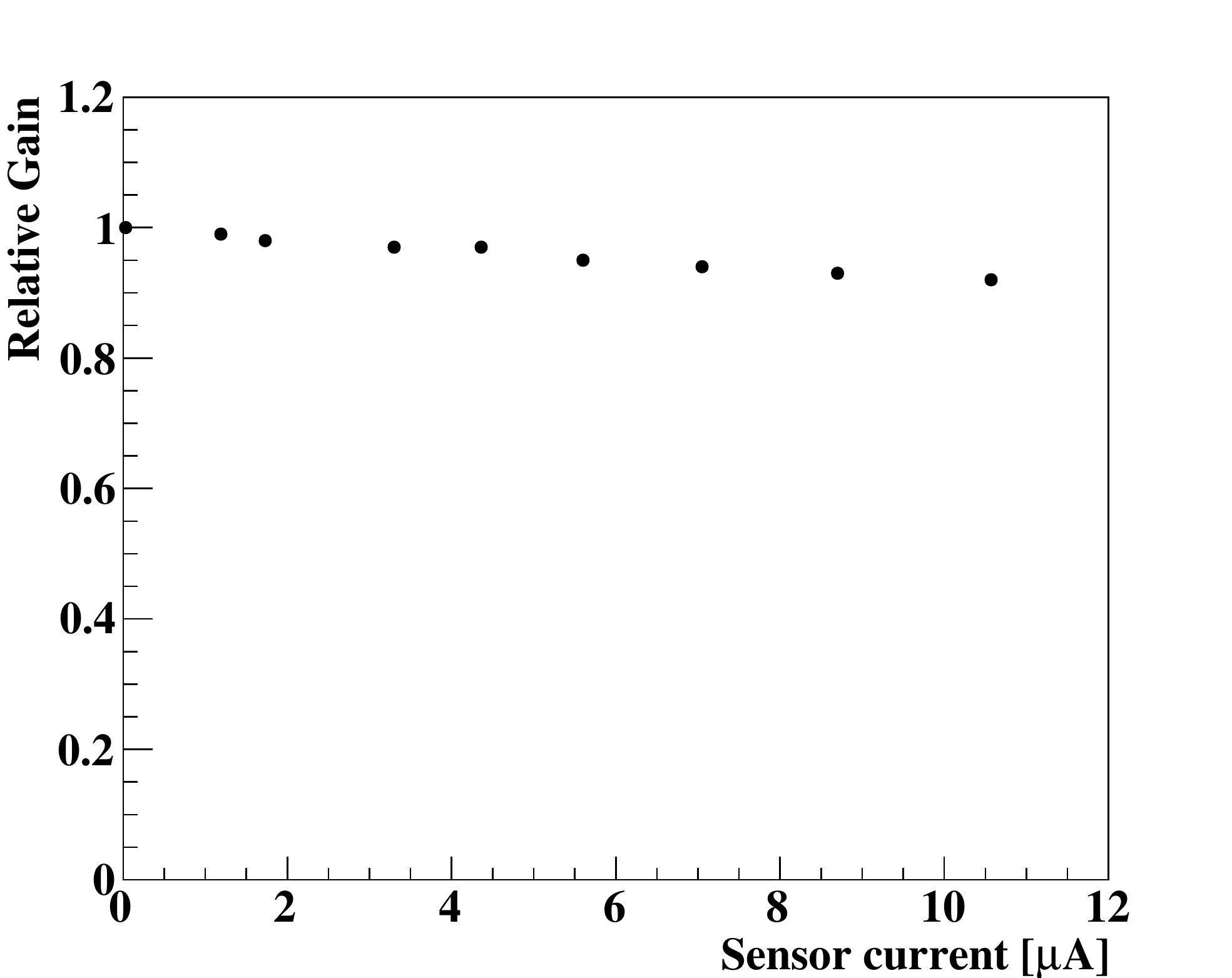}
   \end{center}
   \caption{
  Relative gain measured in LXe as a function of the sensor current increased by 
  background light.
   }
   \label{fig:Gain variation with BG light}
\end{figure}

The potential issue in the scheme of the series connection 
shown in Fig.\,\ref{fig:series connection} is its moderate high-rate performance.
The bias voltage can vary due to the voltage drop in the resistors in the series connection 
and in the protection resistor of the bias circuit 
with the increased sensor current in a high rate environment.
The rate-dependent variation of the bias voltage would cause 
an instability of the sensor gain.
The gain variation was measured in LXe by monitoring the signal charge for light pulses from a blue LED
with an illumination of a background light from another blue LED,
which simulates the background light in a high rate environment.
The gain was measured at different intensities of the background light 
by changing the frequency of the light pulses.
The overall resistance of $23.6\,\mathrm{k}\Omega$ 
(two resistors of $10\,\mathrm{k}\Omega$ 
each in the series connection and one resistor of $3.6\,\mathrm{k}\Omega$ in the bias circuit) 
and a decoupling capacitance of 10\,nF were used in this measurement.
Fig.\,\ref{fig:Gain variation with BG light} shows the relative charge of the primary signal 
as a function of the increase of the sensor current due to the background light,
where the gain variation was measured to be $0.8\%/\mu\mathrm{A}$.
The average increase of the sensor current 
due to the background light expected 
in the MEG II experiment is about $2\,\mu\mathrm{A}$ and thus,
the gain degradation is estimated to be only a few percent.
A larger variation of the background light would, however, be caused 
by an unexpectedly large instability of the muon beam used in the experiment.
A smaller resistance ($2\,\mathrm{k}\Omega$ 
each in the series connection)
is, therefore, adopted in the final version of the readout PCB
as shown in Fig.\,\ref{fig:series connection} to minimise the voltage drop.
The resistance for the final version of the bias circuit is also supposed to be smaller 
although an optimal resistance is still under study because the voltage across the resistance 
is also used to monitor the sensor current.
A larger decoupling capacitance (22\,nF) is used 
to suppress the signal leakage to the smaller resistance.

\subsection{Test of Sensors Produced for MEG II LXe Detector at Room Temperature}

We produced 4200 VUV-MPPCs including 108 spares for the MEG II LXe detector 
and tested 4180 sensors at room temperature.
The test includes measurements of the current-voltage characteristic ($I$-$V$ curve) 
for the individual 16720 ($=4180\times4$) chips and
measurements of the relative signal gain for visible light and the waveform 
for the 4180 sensors.
The breakdown voltage and the dark current were estimated from the $I$-$V$ curve. 
Only a small fraction of the sensors 
(0.2\% of all the chips and 0.8\% of all the sensors)
show a slightly odd behaviour near the breakdown voltage in the $I$-$V$ curve such as a small current offset 
and an unusual shape of the curve 
although most of them can still be operational 
in the LXe detector.

\begin{figure}[htpb]
   \begin{center}
      \includegraphics[width=8cm]{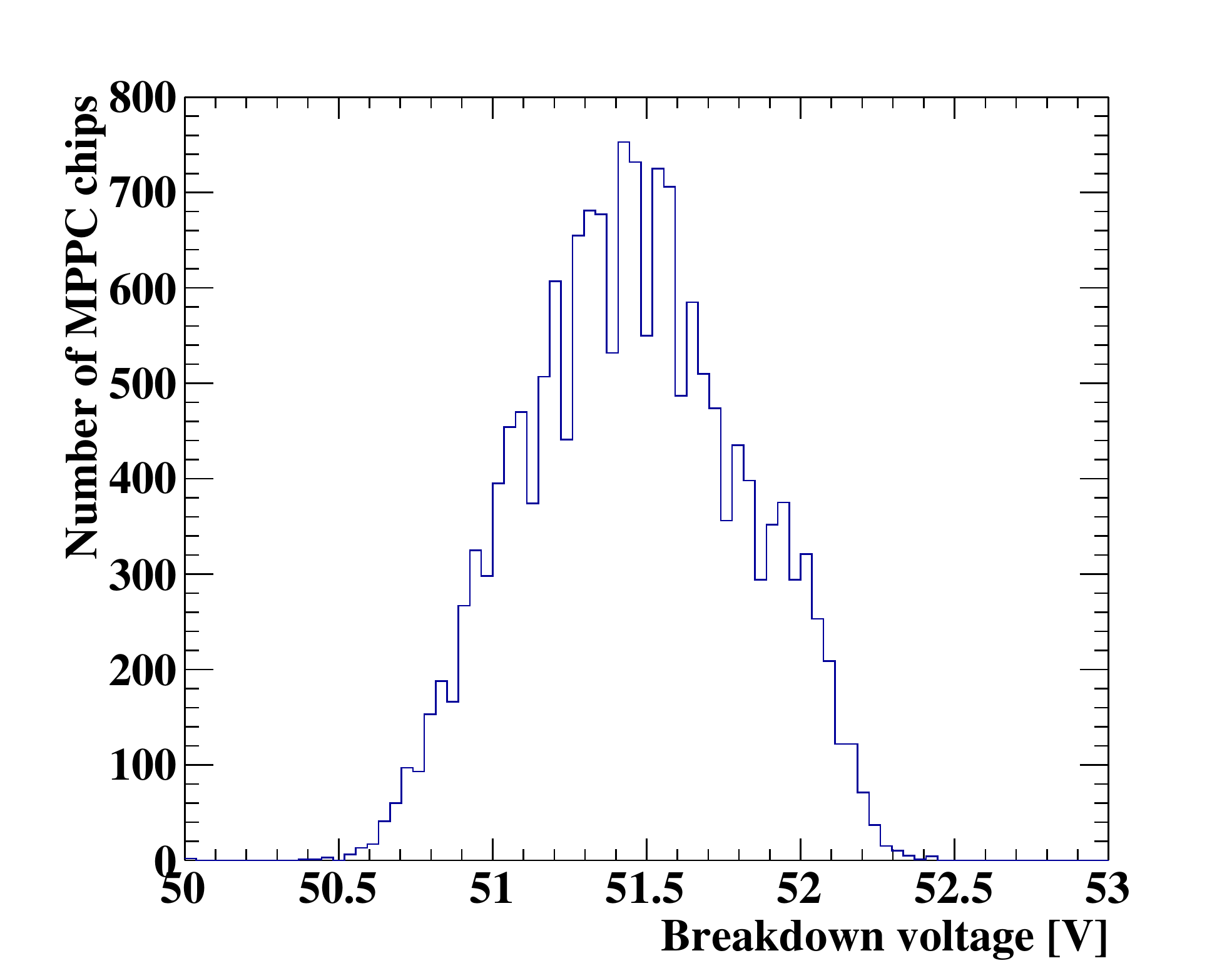}
   \end{center}
   \caption{
  Distribution of the measured breakdown voltage for the 16720 MPPC chips.
   }
   \label{fig:Mass test Vbd}
\end{figure}

\begin{figure}[htpb]
   \begin{center}
      \includegraphics[width=8cm]{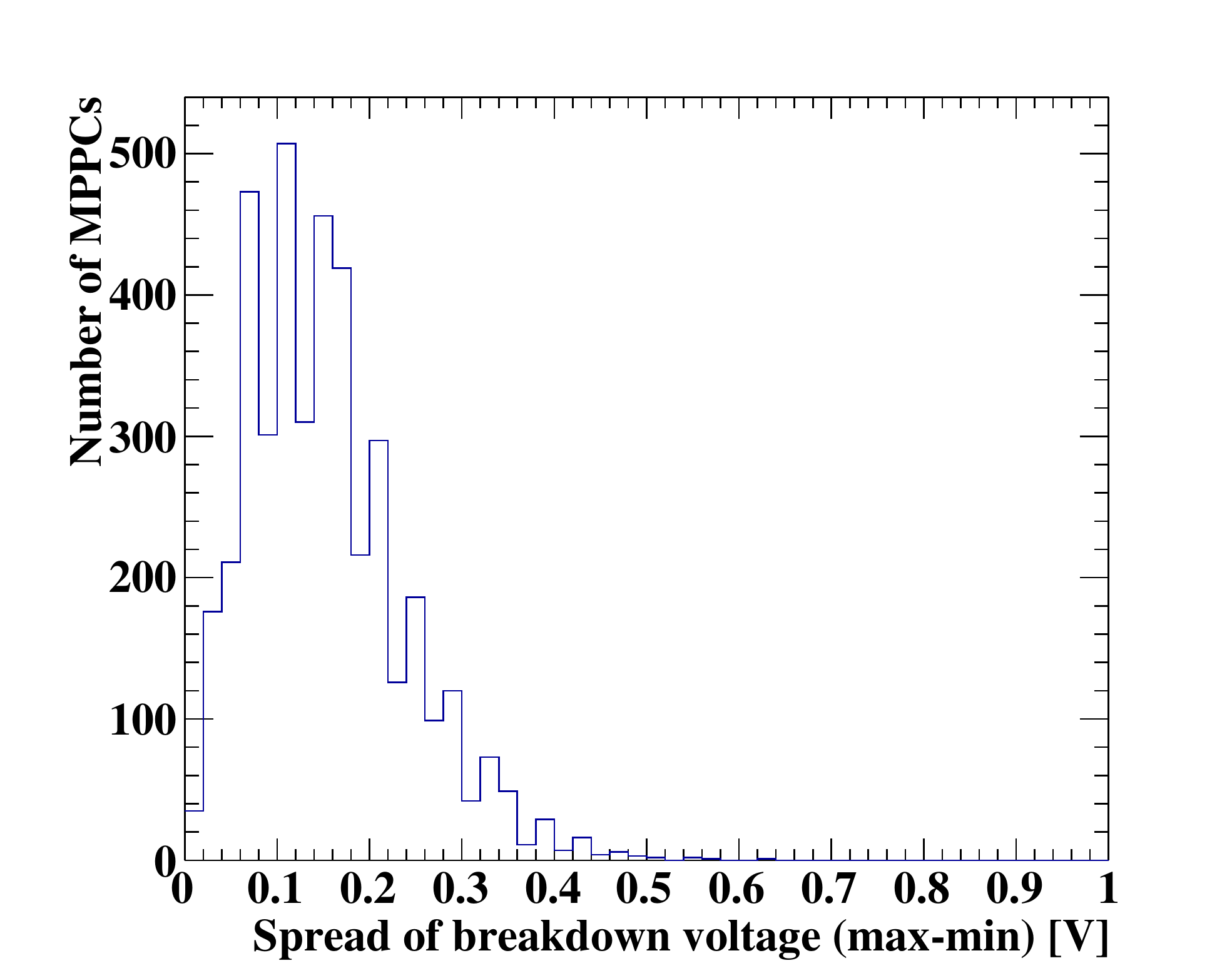}
   \end{center}
   \caption{
  Distribution of the spread of the breakdown voltages 
  for the four MPPC chips within each MPPC.
   }
   \label{fig:Mass test Vbd max-min}
\end{figure}

The breakdown voltage was estimated by fitting the $I$-$V$ curve 
with a linear and a parabolic functions before and after 
the breakdown voltage, respectively.
Fig.\,\ref{fig:Mass test Vbd} shows the distribution of the measured breakdown voltages for the 16720 chips, 
which ranges from 50.5\,V to 52.5\,V.
Although the spread of the breakdown voltages is not very small, 
it is not an issue 
since the four chips with almost the same breakdown voltage 
were selected for the production of each sensor. 
In fact, the spread of the breakdown voltages for the four chips within each sensor, 
which is defined as the difference between the maximum and minimum, was measured to be
only a few hundred mV as shown in Fig.\,\ref{fig:Mass test Vbd max-min}.
This spread of the breakdown voltage within each sensor 
corresponds to a gain variation of only a few percent over the four chips
when they are operated at a common over-voltage of 7\,V.


\begin{figure}[htpb]
   \begin{center}
      \includegraphics[width=8cm]{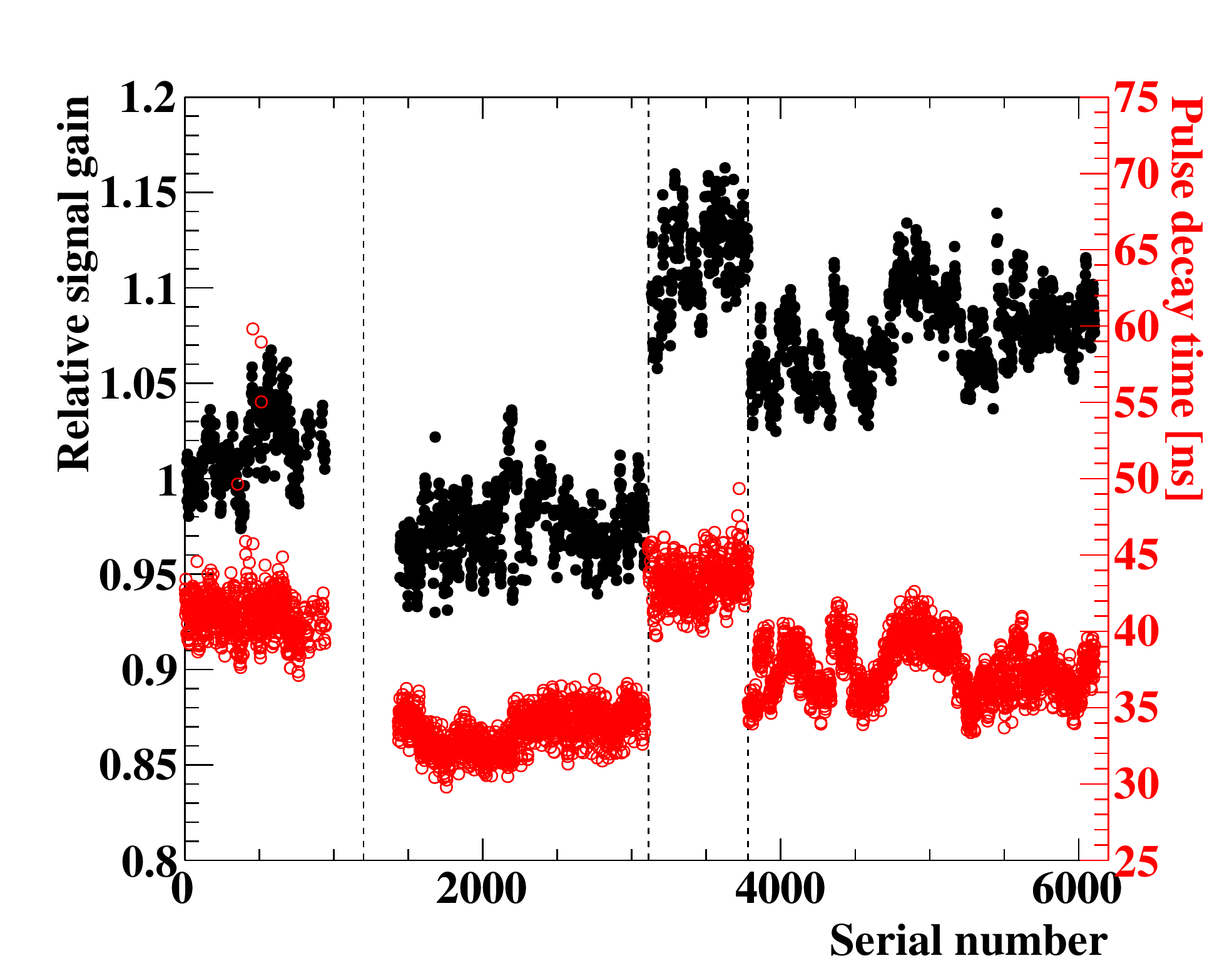}
   \end{center}
   \caption{
  Relative signal gain (black filled circle) and pulse-decay-time (red open circle) measured for all 4180 MPPCs.
   }
   \label{fig:relative gain and pulse decay time at mass test}
\end{figure}

The response to short light pulses from a blue LED was measured
for all the sensors mounted on PCBs with the circuitry 
for the series connection as shown in Fig.\,\ref{fig:series connection}.
The relative signal gain for the LED light pulses was measured by taking a ratio 
of the output charge to that observed for a reference MPPC.
The black filled circle
in Fig.\,\ref{fig:relative gain and pulse decay time at mass test} 
shows the relative signal gain measured at the same over-voltage
as a function of the serial number of the VUV-MPPC 
where a variation of 20\% and a clear dependence on the production lot are seen.
The sensor-by-sensor variation of the signal gain 
is, however, considered to be due mainly to the difference 
in the probabilities of the correlated noises instead of the difference in the PDE.
The pulse-decay-time, which can vary depending on the probability 
of the after-pulsing, was also measured to verify this hypothesis.
In fact the same production lot dependence as in the relative signal gain 
was observed in the measured pulse-decay-time 
as shown as the red open circle
in Fig.\,\ref{fig:relative gain and pulse decay time at mass test}.

\section{Summary and Conclusion}
A large-area VUV-sensitive MPPC has been developed for the LXe scintillation detector
of the MEG II experiment.
The properties of the production model of the VUV-MPPC were thoroughly studied
in a series of measurements in LXe.
It was found to have a high PDE for the LXe scintillation light up to 21\%,
a high gain, low dark count rate and a good single photoelectron resolution.
The overall probability of the correlated noises was also found to be 
sufficiently low,
which enables one to operate it at a high over-voltage up to 7--8\,V.
A large active area of $139\,\mathrm{mm}^2$ was realised 
by connecting four small MPPC chips 
($5.95\times 5.85\,\mathrm{mm}^2$ each) in series 
to avoid the large sensor capacitance and to have a short pulse-decay-time.
Performance tests of 4180 VUV-MPPCs produced 
for the MEG II LXe detector were carried out at room temperature,
where the $I$-$V$ curves were measured for all 16720 chips individually 
and the relative signal gains and the pulse shapes were checked for 4180 sensors.
It turned out that all the sensors with very few exceptions work properly.
The MPPC technology for the enhanced VUV-sensitivity presented here 
can also be applied 
to the detection of the scintillation light 
for other liquid rare-gas such as LAr and LKr. 
In fact, it was reported in Ref.~\cite{Igarashi:2015cma} 
that a MPPC based on the same technology showed a good performance in LAr
to detect the scintillation light from LAr. 
Following our development, another vendor has also started to develop SiPMs with VUV sensitivities~\cite{7182368,nEXO-FBKSiPM}.
The VUV-MPPCs tested here were successfully installed in the MEG II LXe detector, which is now being commissioned.

\section*{Acknowledgements}
This work was supported by JSPS KAKENHI Grant Numbers JP22000004, JP25247034 and JP26000004. K.~Ieki was supported by Grant-in-Aid for JSPS Fellows (Grant Number 2014-0006).      

\section*{References}
\bibliography{mybibfile}

\begin{thebibliography}{10}
\expandafter\ifx\csname url\endcsname\relax
  \def\url#1{\texttt{#1}}\fi
\expandafter\ifx\csname urlprefix\endcsname\relax\def\urlprefix{URL }\fi
\expandafter\ifx\csname href\endcsname\relax
  \def\href#1#2{#2} \def\path#1{#1}\fi

\bibitem{Nakamura:LXeEmissionSpectrum}
K.~Fujii, et~al., {High-accuracy measurement of the emission spectrum of liquid
  xenon in the vacuum ultraviolet region}, Nucl. Instr. and Meth. A 795 (2015)
  293--297.
\newblock \href {http://dx.doi.org/10.1016/j.nima.2015.05.065}
  {\path{doi:10.1016/j.nima.2015.05.065}}.

\bibitem{Aprile:2008bga}
E.~Aprile, et~al., {Noble Gas Detectors}, Wiley, 2008.
\newblock \href {http://dx.doi.org/10.1002/9783527610020}
  {\path{doi:10.1002/9783527610020}}.

\bibitem{MEG-design-paper}
A.~M. Baldini, et~al., {The design of the MEG  II experiment}, Eur. Phys. J. C
  78~(5) (2018) 380.
\newblock \href {http://dx.doi.org/10.1140/epjc/s10052-018-5845-6}
  {\path{doi:10.1140/epjc/s10052-018-5845-6}}.

\bibitem{Ootani:2015cia}
W.~Ootani, et~al., {Development of deep-UV sensitive MPPC for liquid xenon
  scintillation detector}, Nucl. Instr. and Meth. A 787 (2015) 220--223.
\newblock \href {http://dx.doi.org/10.1016/j.nima.2014.12.007}
  {\path{doi:10.1016/j.nima.2014.12.007}}.

\bibitem{NakamuraLXeRefractiveIndex}
S.~Nakamura, et~al., in: {Workshop on Ionization and Scintillation Counters and
  Their Uses (unpublished)}, 2007.

\bibitem{Kowalski-Nuclear-Electronics}
E.~Kowalski, {Nuclear Electronics}, Springer-Verlag, 1970.
\newblock \href {http://dx.doi.org/10.1007/978-3-642-87663-9}
  {\path{doi:10.1007/978-3-642-87663-9}}.

\bibitem{Baldini:2006}
A.~Baldini, et~al., A radioactive point-source lattice for calibrating and
  monitoring the liquid xenon calorimeter of the {MEG} experiment, Nucl. Instr.
  and Meth. A 565 (2006) 589--598.
\newblock \href {http://dx.doi.org/10.1016/j.nima.2006.06.055}
  {\path{doi:10.1016/j.nima.2006.06.055}}.

\bibitem{AcktarBlack}
\href{http://www.acktar.com/category/BlackOpticalCoating}{Acktar
  black{\texttrademark}}.
\newline\urlprefix\url{http://www.acktar.com/category/BlackOpticalCoating}

\bibitem{Ritt:2010zz}
S.~Ritt, et~al., {Application of the DRS chip for fast waveform digitizing},
  Nucl. Instr. and Meth. A 623 (2010) 486--488.
\newblock \href {http://dx.doi.org/10.1016/j.nima.2010.03.045}
  {\path{doi:10.1016/j.nima.2010.03.045}}.

\bibitem{T2K:NIMA2009}
M.~Yokoyama, et~al., {Mass production test of Hamamatsu MPPC for T2K neutrino
  oscillation experiment}, Nucl. Instr. and Meth. A 610 (2009) 362--365.
\newblock \href {http://dx.doi.org/10.1016/j.nima.2009.05.107}
  {\path{doi:10.1016/j.nima.2009.05.107}}.

\bibitem{Doke:1999ku}
T.~Doke, K.~Masuda, {Present status of liquid rare gas scintillation detectors
  and their new application to gamma-ray calorimeters}, Nucl. Instr. and Meth.
  A 420 (1999) 62--80.
\newblock \href {http://dx.doi.org/10.1016/S0168-9002(98)00933-4}
  {\path{doi:10.1016/S0168-9002(98)00933-4}}.

\bibitem{Igarashi:2015cma}
T.~Igarashi, et~al., {Performance of VUV-sensitive MPPC for Liquid Argon
  Scintillation Light}, Nucl. Instr. and Meth. A 833 (2016) 239--244.
\newblock \href {http://arxiv.org/abs/1505.00091} {\path{arXiv:1505.00091}},
  \href {http://dx.doi.org/10.1016/j.nima.2016.07.008}
  {\path{doi:10.1016/j.nima.2016.07.008}}.

\bibitem{7182368}
I.~Ostrovskiy, et~al., Characterization of {S}ilicon {P}hotomultipliers for
  n{EXO}, IEEE Trans. Nucl. Sci. 62~(4) (2015) 1825--1836.
\newblock \href {http://dx.doi.org/10.1109/TNS.2015.2453932}
  {\path{doi:10.1109/TNS.2015.2453932}}.

\bibitem{nEXO-FBKSiPM}
A.~Jamil, et~al., {VUV}-sensitive {S}ilicon {P}hotomultipliers for {X}enon
  {S}cintillation {L}ight {D}etection in n{EXO} (2018).
\newblock \href {http://arxiv.org/abs/1806.02220} {\path{arXiv:1806.02220}}.

\end{thebibliography}

\end{document}